\documentclass[pdflatex,sn-mathphys-num]{sn-jnl}

\usepackage{graphicx}%
\usepackage{multirow}%
\usepackage{amsmath,amssymb,amsfonts}%
\usepackage{amsthm}%
\usepackage[title]{appendix}%
\usepackage{textcomp}%
\usepackage{manyfoot}%
\usepackage{booktabs}%
\usepackage{bookmark}
\usepackage{tikz}%
\usetikzlibrary{arrows.meta}%

\newcommand{\RQ}[1]{\texorpdfstring{\textbf{RQ\textsubscript{#1}}}{RQ#1}}
\newcommand{\finding}[2]{\begin{center}\fbox{\parbox{0.94\textwidth}{\textbf{\RQ{#1} Summary:} #2}}\end{center}}

\begin{document}

\title[Metric-Guided Agentic Refactoring]{From Cosmetic to Structural: A Paired Experiment on Metric-Guided Agentic Refactoring}
%

\author[1]{\fnm{Nadim} \sur{Hagatulah}}\email{nadim.hagatulah@cs.lth.se}

\author[1,2]{\fnm{Markus} \sur{Borg}}\email{markus.borg@codescene.com}

\author[1]{\fnm{Emma} \sur{S\"{o}derberg}}\email{emma.soderberg@cs.lth.se}

\affil[1]{\orgname{Lund University}, \orgaddress{\city{Lund}, \country{Sweden}}}

\affil[2]{\orgname{CodeScene}, \orgaddress{\city{Malm\"{o}}, \country{Sweden}}}

\abstract{The role of coding agents is expanding beyond writing new code toward maintaining existing systems as AI teammates. Yet recent studies show that agent refactorings favor superficial, consistency-oriented edits over structural transformations needed to resolve Technical Debt (TD). We investigate if explicit static-analysis feedback closes this gap. Through the CodeScene MCP server, we give an open-source agent (\texttt{gptme}) access to code smell reports and CodeHealth (CH) scores (1--10, higher is better), and compare guided against unguided refactorings in a paired experiment covering $24{,}620$ single-file competitive programming solutions in C++, Java, and Python. We find that guidance shifts the agent from performing cosmetic edits toward structural operations, raising the rate of effective refactorings by $21$--$26$ percentage points. The benefit depends on initial code quality, with language-dependent break-even points at CH $4.9$--$7.1$, below where the structural transformations pushed by guidance increasingly risk regression, indicating where the agent needs more support to refactor. Guided runs also shift the failure profile toward logical test failures, introduce new coupling smells, and consume $\sim$$3$--$5\times$ more tokens over $\sim$$2\times$ as many interaction turns. Our results indicate that AI teammates can benefit from similar static-analysis feedback that human developers already use, enabling structural transformations toward more effective TD remediation, at a higher token cost and more interaction turns.
}


\maketitle

\section{Introduction}
\label{sec:intro}

Refactoring is the established practice for remediating code smells~\citep{fowler_refactoring_2018}, which accumulate as Technical Debt (TD) when left unresolved. TD has been found to lead to inflated development costs~\citep{moserCaseStudyImpact2008, avgeriou_2023_td_management}, increased defect and security risks~\cite{tornhillCodeRedBusiness2022, Siavvas2020}, while also increasing developer unhappiness~\citep{GRAZIOTIN201832, BESKER201941}. Consequently, managing TD is important for sustainable, long-term software development~\citep{Biazotto_2025_insights}.

Despite the positive effects of refactoring~\citep{refactoringMicrosoft, refactoringProgramComprehension, moserCaseStudyImpact2008}, it remains an infrequent activity among developers~\citep{howWeRefactor2012}. \citet{martini_technical_2018} report that only 26\% of developers use tools to track TD, and 7\% of organizations systematically monitor it. One reason is that refactoring requires domain knowledge and a deep understanding of the existing code to anticipate the consequences of changes~\citep{chen_2016_perspective_on_refactoring}. Moreover, improper refactoring risks altering semantic behavior or introducing bugs~\citep{rejectRefactorings2021}. The activity is also frequently deprioritized in favor of other tasks such as feature development~\citep{ace2025Tornhill}. As a result, developers often leave design issues unresolved rather than risk breaking functionality.

Large Language Models (LLMs) offer an automated alternative, generating refactored versions of code on demand~\citep{cordeiroLLMdrivenCodeRefactoring2025, chatgptUsageInOpenSource2024, em-assist2024, shirafujiRefactoringProgramsUsing2023, ace2025Tornhill}. The advancement of LLMs has further enabled autonomous multi-turn agents that plan, act through tools, and iterate on complex Software Engineering (SE) tasks with reduced human intervention~\citep{wang2025aiagenticprogrammingsurvey}. This signals a shift toward what \citet{li2025riseaiteammatessoftware} describe as AI teammates in SE. If such agents are to act as teammates on projects, they must also take on code maintenance.

Large-scale studies of agentic refactoring by \citet{horikawa2025agenticrefactoringempiricalstudy} and \citet{ottenhof2026agentsrefactorempiricalstudy} show that agent-authored refactoring commits are dominated by superficial, consistency-oriented edits (e.g., variable renaming and annotation changes) rather than the structural transformations that remediate design flaws (e.g., Extract Method). \citet{cordeiroLLMdrivenCodeRefactoring2025} attribute such misaligned modifications to a missing explicit purpose. Without concrete refactoring goals, LLMs fall back on their latent notion of clean code, and the authors suggest enriching prompts with guided information about specific issues. Furthermore, on the concept of AI-Friendliness, introduced from Thoughtworks Technology Radar~\cite{thoughtworks2025}, \citet{borg2026codemachinesjusthumans} empirically investigated this direction, showing that code written for human comprehension is also easier for LLMs to refactor without breaking semantic behavior.

These observations raise the question of what happens when an autonomous coding agent is given access to the same static-analysis feedback that modern development infrastructure already provides. Quality gates in IDEs and continuous integration (CI) pipelines routinely surface maintainability issues to human developers\footnote{\url{https://www.sonarsource.com/products/sonarqube/}}\footnote{\url{https://codescene.com/product/automated-code-health-reviews}}, and the Model Context Protocol (MCP) makes it practical to expose the same tooling to agents.

In this study, we evaluate the impact of providing an agent with explicit static-analysis feedback through the CodeScene MCP Server\footnote{\url{https://github.com/codescene-oss/codescene-mcp-server}} (version 0.1.2, hereafter, the MCP server), exposing tools that report the occurrence, severity, and position of code smells, together with an aggregated CodeHealth (CH) score of the file. The CH metric has been shown to predict defect count and development velocity~\citep{borgIncreasingNotDiminishing2024, tornhillCodeRedBusiness2022, borg_ghost_2024}. We run a paired experiment on $24{,}620$ files across C++, Java, and Python ($49{,}240$ agent runs), using the open-source \texttt{gptme} agent harness powered by the open-weight Qwen3-Coder-30B-A3B model, comparing a \emph{Basic} instruction (the agent relies on its intrinsic judgment) against a \emph{Guided} instruction (the agent iterates with the MCP tool). Throughout the paper, \emph{metric guidance} refers to the Guided configuration as a whole, which bundles the metric-referencing instruction, the availability of the tool, and the grounding of each iteration in external tool feedback (Section~\ref{sec:setup}).

We evaluate the agents on single-file competitive programming solutions instead of production modules. While production codebases offer realism, they are more complex, often lacking complete test suites, controlled quality distributions, and reproducibility required for a strict, paired experimental design. By trading external realism for internal validity, our corpus isolates the agent's structural reasoning from complexities in enterprise, such as cross-file dependencies and build systems. Consequently, our findings measure how metric feedback alters agent behavior, rather than projecting end-to-end maintenance performance on production codebases.

The study is structured around the following Research Questions (RQs), moving from what guidance achieves, through what the agent does differently, to when it helps and what the overhead is:

\begin{itemize}
    \item \RQ{1}\textbf{: How does metric guidance affect the outcomes of agentic refactoring?} \\
    We compare the pass rate (proxy for behaviour preservation), CH improvement rate, and the combined effective refactoring rate between Basic and Guided runs to measure how guidance changes what the agent achieves.
    \item \RQ{2}\textbf{: How does metric guidance change the refactoring operations applied and the code smells addressed?} \\
 The outcome does not necessarily distinguish structural work from superficial edits. We check the applied refactoring operations and per-smell remediation rates to characterize the work of each agent.
    \item \RQ{3}\textbf{: How does the initial CH of a file moderate the effect of metric guidance?} \\
    Motivated by the AI-Friendliness hypothesis~\citep{borg2026codemachinesjusthumans}, we test if the potential benefit of guidance depends on how degraded the code is to begin with.
    \item \RQ{4}\textbf{: How does metric guidance change the failure modes of agentic refactoring?} \\
    When guided refactorings fail, do they fail differently? We classify failures into compile errors, runtime errors, test failures, and exceptions, and map error messages to their causes.
    \item \RQ{5}\textbf{: What computational overhead does metric-guided refactoring introduce?} \\
    We expect a tool-grounded feedback loop to bring increased tokens and interaction turns resulting in a trade-off that practitioners need to consider.
\end{itemize}

We present a large-scale, paired evaluation of static-analysis guidance for an agent performing refactoring operations, covering three languages, $24{,}620$ files, and $49{,}240$ agent runs. We provide evidence that metric guidance redirects agent behavior from superficial refactorings toward more meaningful structural transformations, raising the effective refactoring rate by $21$ to $26$ percentage points (pp) under the guiding metric, while also introducing some new coupling-related trade-off smells. Our moderation analysis extends the AI-Friendliness hypothesis to the agentic setting, finding that the effect of metric guided refactorings grows with the initial CH of the file, with language-specific break-even points, the initial CH at which Guided and Basic runs are equally likely to produce an effective refactoring. It ranges from CH $4.9$ (C++) to $7.1$ (Python). Above it, guidance helps by up to $+34$ pp, while below it, guidance lowers the effective refactoring rate (driven by a lower test pass rate). These break-even points are specific to our metric, LLM, harness, and languages, and should not be read as general CH thresholds for agentic refactoring. We further characterize how guidance shifts the failure profile toward test failures, and quantify the overhead of the iterative feedback loop ($\sim$$2\times$ turns, $3$--$5\times$ tokens). All data, prompts, and analysis scripts are available in a replication package~\citep{replication}.

\section{Related Work}
\label{sec:related}

Automated refactoring support ranges from rule-based tools~\citep{jextract2014, jdeoderant2011}, and machine-learning classifiers~\citep{REMS2023, GEMS2017}, to LLM-based approaches~\citep{em-assist2024, ace2025Tornhill}, and multi-agent frameworks~\citep{MUARFMultiAgentRefactoring, karabiyikRefactorGPTChatGPTbasedMultiagent2025, haskellRefactoringMultiAgent, pythonMultiAgentRef, RefAgentMultiagentLLMbased2025}. While they are capable, they require human triggers, supervision, and intervention. LLM-based single-shot approaches, in turn, operate on isolated snippets and cannot verify the effect of their changes. Autonomous agents, in contrast, read files, apply edits, invoke tools, and decide when they are done~\citep{wang2025aiagenticprogrammingsurvey}.

Recent large-scale empirical evaluations have compared the refactoring behavior of autonomous agents against human-driven maintenance. \citet{horikawa2025agenticrefactoringempiricalstudy} demonstrated that refactoring is a common activity in agentic development, but it is skewed toward low-level consistency-oriented changes (e.g., renaming variables and parameters or changing variable types), in contrast to higher-level structural changes that are more common in human refactoring, highlighting a drift in refactoring intent. Agents tend to prefer localized improvements. Similarly, \citet{ottenhof2026agentsrefactorempiricalstudy} observed that agents frequently perform annotation changes instead of structural modifications, often failing to fix underlying code smells and design issues.

A recent empirical study of 567 GitHub pull requests generated autonomously by Claude Code revealed that while developers frequently deploy agents for refactoring tasks, 45.1\% of the merged agentic pull requests still required subsequent human revisions~\citep{Wanatabe2026}. This further shows that, while agents can propose extensive changes, they struggle to finalize structural design and adhere to project-specific standards without explicit guidance.

A candidate explanation is missing purpose. \citet{cordeiroLLMdrivenCodeRefactoring2025} highlight that LLMs operating without explicit goals misalign with the refactoring intent of improving maintainability, and argue for enriching prompts with guided information about the specific issues to address. Our study operationalizes this suggestion in an agentic setting. Instead of statically enriching a prompt, we give the agent a tool to query iteratively.

CH is a static source-code metric that aggregates code smells and their severity into a score from 1 (severely degraded) to 10 (perfectly healthy). CH has been validated as a predictor of TD consequences, where low-CH files carry higher defect volume and slower development velocity~\citep{tornhillCodeRedBusiness2022, borgIncreasingNotDiminishing2024, borg_ghost_2024}.

\citet{borg2026codemachinesjusthumans} empirically investigated AI-Friendliness~\cite{thoughtworks2025}, demonstrating that code written for human comprehension is also more tractable for AI. High initial internal code quality predicts successful AI-generated refactorings and significantly lower break rates. Their study evaluated single-shot refactoring of Python files. We extend this line of work to multi-turn agents and two additional languages, and we examine whether giving the agent the metric itself as an interactive tool changes the outcome.

We still lack empirical evidence on how effectively agents use metric-guided feedback to fix code smells. It is unknown (i) if agents can successfully interpret and act on smell reports for structural transformations, (ii) how such guidance interacts with the initial quality of the code, and (iii) what the guidance introduces as overhead in correctness risk and compute. Our study addresses this gap by connecting an agent to a static-analysis tool via MCP and performing a large-scale, multi-language, paired evaluation of multi-turn, metric-guided refactoring.

\section{Methodology}
\label{sec:method}

This section presents the construction of the dataset, selection of the agent and model, the experimental setup comparing baseline and metric-guided instructions, the evaluation metrics, and the statistical analyses that address the RQs.

\subsection{Dataset Construction}
\label{sec:dataset}

The dataset is derived from the CodeContests dataset~\citep{code_constest_2022}, which contains $\sim$$12{,}000{,}000$ solutions to competitive programming tasks. The final experimental dataset contains three subsets: C++ ($9{,}825$ solutions), Java ($9{,}795$ solutions), and Python ($5{,}000$ solutions). Each solution comes with STDIN/STDOUT test suites, providing a behavioral oracle for judging if a refactoring preserved semantic behavior. The Python subset is reused from the replication package of \citet{borg2026codemachinesjusthumans}.

The C++ and Java subsets are constructed with a stratified method: (i) randomly pick $2{,}500$ solutions with CH $\ge 9$, (ii) randomly pick $2{,}500$ solutions with CH $< 9$, (iii) sort all solutions by CH score and add the lowest-scoring ones until $10{,}000$ total solutions have been selected, and (iv) merge and de-duplicate the sets from the previous steps (dropping $175$ solutions from the C++ subset and $205$ from the Java subset). For each sampling stage (i--iii), CodeBLEU~\citep{ren2020codebleu} is applied as a syntactic similarity filter (weights $n$-gram $=0.1$, weighted-$n$-gram $=0.4$, AST-match $=0.5$, dataflow-match $=0$). It acts as a max-similarity gate. For each candidate solution, the CodeBLEU score is computed against all already-selected solutions, and the candidate is accepted only if the maximum similarity is below $0.90$, encouraging a diversely sampled dataset. The first two stages, (i) and (ii), are identical to \citet{borg2026codemachinesjusthumans} dataset construction. The third stage extends the dataset with low-CH solutions to cover a wider span of initial code quality solutions, which \RQ{3} requires.

Table~\ref{tab:dataset} summarizes the resulting subsets. The C++ and Java subsets span a wide range of initial CH, while the Python subset, reused as-is from previous work, is concentrated in the healthy range with only $10$ solutions below CH~6. The CodeContests dataset contains substantially fewer Python solutions, thus the same approach cannot be applied for that language.

\begin{table}[htbp]
    \centering
    \caption{Descriptive statistics of the dataset. Initial CH and Source Lines of Code (SLoC) are computed on the original files.}
    \label{tab:dataset}
    \setlength{\tabcolsep}{4pt}
    \begin{tabular}{@{} l r c c r c c c @{}}
        \toprule
        & & \multicolumn{3}{c}{\textbf{Initial CH}} & \multicolumn{3}{c}{\textbf{SLoC}} \\
        \cmidrule(lr){3-5} \cmidrule(lr){6-8}
        \textbf{Language} & \textbf{Solutions} & \textbf{Mean $\pm$ SD} & \textbf{Median} & \textbf{$<6$ ($N$, \%)} & \textbf{Mean $\pm$ SD} & \textbf{Median} & \textbf{Max} \\
        \midrule
        C++    & $9{,}825$ & $7.80 \pm 1.26$ & $7.3$ & $479$ ($4.9$) & $135 \pm 33$ & $129$ & $290$ \\
        Java   & $9{,}795$ & $8.07 \pm 1.09$ & $7.7$ & $237$ ($2.4$) & $167 \pm 39$ & $167$ & $299$ \\
        Python & $5{,}000$ & $9.02 \pm 0.65$ & $9.0$ & $10$ ($0.2$)  & $109 \pm 40$ & $100$ & $537$ \\
        \bottomrule
    \end{tabular}
\end{table}

\subsection{Agent and Model Selection}
\label{sec:agent}

We selected \texttt{gptme}\footnote{\url{https://github.com/gptme/gptme}} (version 0.31.0) as the autonomous CLI coding agent harness. Unlike vendor-locked alternatives such as Claude Code\footnote{\url{https://github.com/anthropics/claude-code}}, Codex\footnote{\url{https://github.com/openai/codex}}, or Agy CLI\footnote{\url{https://github.com/google-antigravity/antigravity-cli}}, \texttt{gptme} is open-source and model-agnostic, enabling integration of locally hosted models. It provides the file-manipulation tools required by our methodology (read, overwrite, patch) and supports MCP for tooling extensions.

The underlying LLM is Qwen3-Coder-30B-A3B-Instruct\footnote{\url{https://huggingface.co/Qwen/Qwen3-Coder-30B-A3B-Instruct}}~\citep{qwen3technicalreport}, a state-of-the-art open-weight model in its size class that fits in the resources of a single NVIDIA A100 (80GB VRAM) GPU. This LLMs Mixture-of-Experts architecture enables efficient, low-latency experiment runs~\citep{artetxe_efficient_2022}. Sampling parameters follow the model card recommendations, with temperature $=0.7$, top\_p $=0.8$, top\_k $=20$, and repetition penalty $=1.05$. The context length is set to $32{,}768$ tokens, and the model is served with vLLM (version 0.15.0)~\citep{kwon_efficient_2023}. We use  an open harness and open weights to keep the full experimental stack reproducible.

\subsection{Experimental Setup}
\label{sec:setup}

\begin{figure}[htbp]
    \centering
    \definecolor{basiccol}{HTML}{B0642A}
\definecolor{guidedcol}{HTML}{2A6BB0}
\begin{tikzpicture}[
    x=1cm, y=1cm, font=\fontsize{6.5}{7.8}\selectfont, >={Stealth[length=1.4mm, width=1.1mm]},
    panel/.style={draw=black!55, fill=black!3, rounded corners=2pt, line width=0.5pt},
    stage/.style={draw=black!55, fill=white, rounded corners=1.5pt, line width=0.4pt, align=center,
                  minimum width=1.3cm, minimum height=0.5cm, inner sep=1.2pt},
    arm/.style={rounded corners=1.5pt, line width=0.6pt, align=center,
                minimum width=1.35cm, minimum height=0.56cm, inner sep=1.2pt},
    flow/.style={->, line width=0.45pt, black!70},
    back/.style={flow, out=90, in=90, looseness=0.55},
    note/.style={font=\fontsize{6}{7}\selectfont\itshape, black!70, inner sep=1pt},
    dot/.style={circle, fill=black!75, inner sep=0pt, minimum size=4.2pt},
    moderator/.style={circle, draw=black!75, line width=0.7pt, inner sep=0pt, minimum size=4.2pt},
]

\node[panel, anchor=south west, minimum width=3.2cm, minimum height=3.6cm] at (0, 0) {};
\node[anchor=north, align=center] at (1.6, 3.5) {%
    \textbf{\fontsize{7.5}{9}\selectfont Dataset}\\[4pt]
    CodeContests solutions\\ stratified by initial CH,\\ CodeBLEU-deduplicated\\[5pt]
    \begin{tabular}{@{}l@{\hspace{1.5em}}r@{}}
        C++    & 9,825 \\
        Java   & 9,795 \\
        Python & 5,000 \\
    \end{tabular}\\[5pt]
    STDIN/STDOUT tests\\ per solution, hidden\\ from the agent};

\node[panel, anchor=south west, minimum width=9.6cm, minimum height=3.6cm] at (3.4, 0) {};
\node[anchor=north west, align=left] at (3.5, 3.5) {%
    \textbf{\fontsize{7.5}{9}\selectfont Paired refactoring runs}\enspace 2 $\times$ 24,620 $=$ 49,240\\[1pt]
    gptme + Qwen3-Coder-30B-A3B, one isolated file per run};

\node[arm, draw=basiccol, fill=basiccol!12] at (4.25, 2.0) {\textbf{Basic}\\self-assessed};
\node[stage] (b1) at (5.85, 2.0) {read\\file};
\node[stage] (b2) at (7.45, 2.0) {identify\\issues};
\node[stage] (b3) at (9.05, 2.0) {patch\\file};
\node[stage] (b4) at (10.65, 2.0) {re-read\\file};
\node[stage, minimum width=0.8cm] (b5) at (12.35, 2.0) {stop};
\draw[flow] (b1) -- (b2);
\draw[flow] (b2) -- (b3);
\draw[flow] (b3) -- (b4);
\draw[flow] (b4) -- (b5);
\draw[back] (b4.north) to node[note, above] {issues left} (b3.north);

\node[arm, draw=guidedcol, fill=guidedcol!12] at (4.25, 0.95) {\textbf{Guided}\\tool feedback};
\node[stage] (g1) at (5.85, 0.95) {read\\file};
\node[stage, draw=guidedcol] (g2) at (7.45, 0.95) {review\\(MCP tool)};
\node[stage] (g3) at (9.05, 0.95) {patch\\file};
\node[stage, draw=guidedcol] (g4) at (10.65, 0.95) {re-review\\(MCP tool)};
\node[stage, minimum width=0.8cm] (g5) at (12.35, 0.95) {stop};
\draw[flow] (g1) -- (g2);
\draw[flow] (g2) -- (g3);
\draw[flow] (g3) -- (g4);
\draw[flow] (g4) -- (g5);
\draw[back] (g4.north) to node[note, above] {smells left} (g3.north);
\node[draw=guidedcol, fill=guidedcol!8, rounded corners=1.5pt, line width=0.4pt, inner sep=1.5pt,
      minimum width=4.0cm, minimum height=0.34cm] (mcp) at (9.05, 0.27)
    {CodeScene MCP server: smells (type, location), CH};
\draw[<->, line width=0.45pt, guidedcol] (g2.south) -- (g2.south |- mcp.north);
\draw[<->, line width=0.45pt, guidedcol] (g4.south) -- (g4.south |- mcp.north);

\draw[line width=0.6pt] (0, -0.7) -- (13.0, -0.7);
\draw[flow] (1.6, 0) -- node[note, right] {original file, tests} (1.6, -0.7);
\draw[flow] (8.2, 0) -- node[note, right] {refactored file, session statistics} (8.2, -0.7);

\node[anchor=west] at (0.05, -1.03) {\textbf{Instrument}};
\node[anchor=west] at (2.3, -1.03) {\textbf{Measure (per run)}};
\foreach \c/\name in {1/Outcomes, 2/Operations, 3/Moderation, 4/Failures, 5/Cost}
    \node[align=center] at ({5.185 + 1.43*\c}, -1.03) {\RQ{\c}\\\name};
\draw[line width=0.4pt] (0, -1.38) -- (13.0, -1.38);

\fill[black!4] (0, -1.38) rectangle (13.0, -2.1);
\fill[black!4] (0, -3.18) rectangle (13.0, -3.54);
\fill[black!4] (0, -3.9) rectangle (13.0, -4.26);
\foreach \r/\group in {1.5/Test oracle, 4/CodeScene CLI, 6/Derived, 7/RefactoringMiner, 8/diff, 9/gptme}
    \node[anchor=west, font=\fontsize{6.5}{7.8}\selectfont\itshape, black!70] at (0.05, {-1.2 - 0.36*\r}) {\group};
\foreach \r/\measure in {
        1/{Pass (compiles, tests pass)},
        2/{Failure stage, error cause},
        3/{Initial CH},
        4/{Final CH, CH improved},
        5/{Smells before and after},
        6/{\textbf{Effective} (pass $\wedge$ CH$\uparrow$)},
        7/{Operation levels (Java, Python)},
        8/{Churn (changed lines)},
        9/{Tokens, interaction turns}}
    \node[anchor=west] at (2.3, {-1.2 - 0.36*\r}) {\measure};
\foreach \r/\c in {1/1, 1/3, 1/4, 2/4, 4/1, 4/3, 5/2, 6/1, 6/3, 7/2, 8/4, 9/5}
    \node[dot] at ({5.185 + 1.43*\c}, {-1.2 - 0.36*\r}) {};
\node[moderator] at ({5.185 + 1.43*3}, {-1.2 - 0.36*3}) {};

\draw[line width=0.4pt] (0, -4.62) -- (13.0, -4.62);
\node[anchor=west] at (0.05, -4.95) {\textbf{Analysis}};
\foreach \c/\test in {1/{McNemar,\\Wilcoxon}, 2/{Wilcoxon,\\McNemar}, 3/{GEE\\(CH$^*$)},
                      4/{permutation\\test}, 5/{Wilcoxon}}
    \node[align=center] at ({5.185 + 1.43*\c}, -4.95) {\test};
\draw[line width=0.6pt] (0, -5.28) -- (13.0, -5.28);
\end{tikzpicture}
    \caption{Experimental design. Each solution is refactored once under each condition (Basic/Guided), with the same agent, model, and environment. Only the instruction and the available tool differ. The table maps the data collected per run to the RQs ($\bullet$ outcome, $\circ$ moderator) and lists the statistical analysis of each RQ.}
    \label{fig:design}
\end{figure}
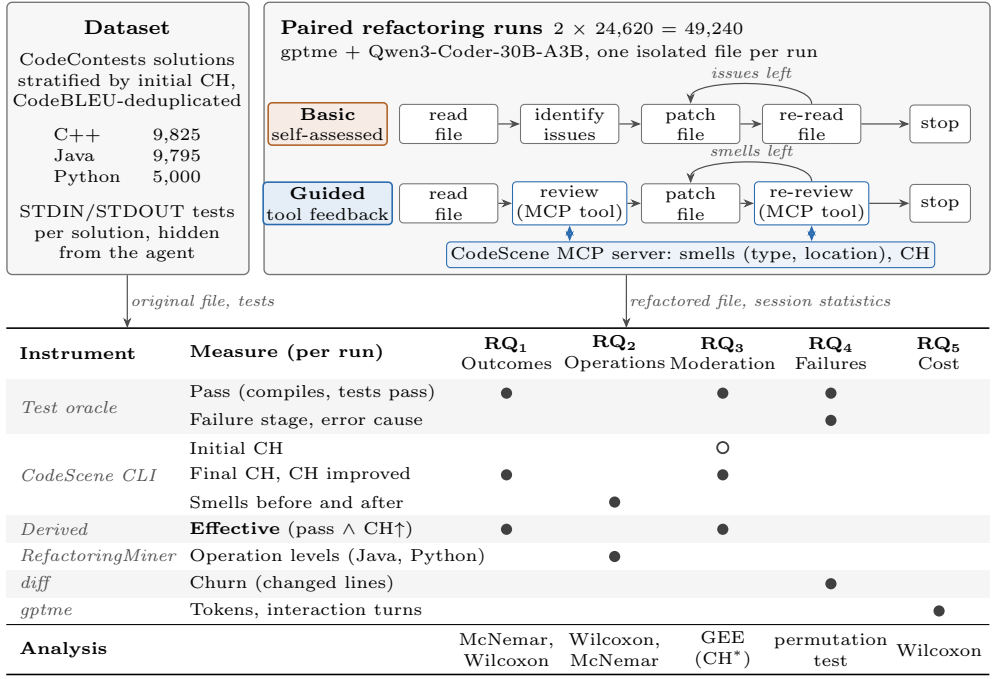

Figure~\ref{fig:design} gives an overview of the experimental design. For each solution in the dataset, we establish an isolated environment by placing the source file in its own dedicated directory. The agent only has visibility of the source file to refactor. In particular, it has no access to the test suite, purposefully simulating maintenance under poor test coverage. We then start a non-interactive, no-confirm instance of \texttt{gptme} in each directory, tasking the agent with refactoring the code to improve maintainability. Each solution is refactored twice under the different conditions (Basic/Guided), yielding a paired design:

\begin{itemize}
    \item \textbf{Basic}: The agent is instructed to improve maintainability by (i) reading the file, (ii) identifying maintainability issues, relying on the LLMs intrinsic capabilities without external tools, (iii) remediating the identified issues and overwriting the file, and (iv) re-reading the file. If unresolved or new issues are found, loop back to step (iii). Otherwise, terminate.
    \item \textbf{Guided}: The agent has the same objective but additionally has access to the CodeScene MCP server, which exposes the \texttt{code\_health\_review} tool: (i) read the file, (ii) identify code smells by invoking \texttt{code\_health\_review}, (iii) remediate the identified issues and overwrite the file, and (iv) re-invoke \texttt{code\_health\_review}. If the tool still reports code smells, loop back to step (iii). Otherwise, terminate.
\end{itemize}

The MCP server is only exposed under the Guided instruction. Both instructions define an explicit goal of improving maintainability, but encourage no specific editing. Both encourage an iterative multi-turn approach, while the execution strategy and the decision to terminate are governed by the agent's own judgment. The only experimental differences between Basic and Guided are the instruction text and the available tools. The model, runtime, and environment are held constant. The exact prompts are listed in Appendix~\ref{sec:prompt-templates}.

The CodeScene MCP server registers thirteen tools. The Guided loop is driven by \texttt{code\_health\_review}, which reports a file's code smells (category, severity, and location) together with its CH score. The agent may also call two auxiliary tools: \texttt{code\_health\_score}, which returns the numeric CH score alone, and \texttt{explain\_code\_health}, which returns a static description of how CH is computed and interpreted rather than any file-specific analysis. The remaining ten tools serve repository- and organization-level workflows, such as technical-debt hotspots, code ownership, pre-commit check, and refactoring business cases, that do not apply to our isolated single-file refactoring scenarios.

The Guided treatment is a bundle, containing the metric-referencing instruction, the availability of the tool, and the external feedback that grounds each iteration. Both instructions encourage iterating, but only the Guided agent iterates against tool-verified findings rather than its own self-assessment. Throughout the paper, ``metric guidance'' is shorthand for this configuration as a whole. Our design compares the bundle against the Basic baseline and does not isolate the causal contribution of the metric content from that of having any iterative feedback loop.

\subsection{Evaluation Metrics}
\label{sec:metrics}

Figure~\ref{fig:design} maps each metric to RQ. We track the following metrics:

\begin{itemize}
    \item \textbf{Pass Rate (\%)}: Rate of refactored solutions that compile (g++ $13.3.0$, javac $24.0.2$, CPython $3.13.12$) and pass the full provided STDIN/STDOUT test suite.
    \item \textbf{CodeHealth (CH, 1--10)}: Computed with the CodeScene CLI (v1.0.17, default configuration) per file in isolation. The same metric and version are used for the Guided instruction and for evaluation.
    \item \textbf{Improved CH Rate (\%)}: Rate of refactored solutions whose CH score strictly increases over the initial file, regardless of the solution still compiling or passing test cases.
    \item \textbf{Effective Refactoring Rate (\%)}: The primary success criterion. A refactoring is considered effective only if it (i) compiles, (ii) passes the test suite, and (iii) strictly increases the CH score. Consequently, since the Guided agent explicitly targets this CH score during its run, we interpret success here as an improvement under CodeScene's specific rules of CH as a proxy for maintainability, rather than as a general claim of full objective maintainability.
    \item \textbf{Interaction Turns}: The number of assistant messages in the conversation, as counted by the \texttt{gptme} session statistics. Each assistant message is one model response, which may contain reasoning, code edits, and tool invocations. User instructions, tool outputs, and system messages are not counted.
    \item \textbf{Token Usage}: $T_{total} = T_{in} + T_{out}$, counted with the \texttt{tiktoken} \texttt{cl100k\_base} tokenizer\footnote{\url{https://github.com/openai/tiktoken}}, reported from \texttt{gptme} session statistics, over the tokens delivered to and generated by the model.
\end{itemize} 

In $752$ refactorings, the CodeScene CLI returns no CH score. In these cases, the agent wrote no refactored file, either because it failed with an error, timed out, or terminated before finishing. Runs without a refactored file count as failed and are classified as Exceptions (Section~\ref{sec:failure-taxonomy}). Runs without a CH score count as not improved, and the CH score, smell list, token count, and turn count comparisons use only the cases where both Guided and Basic runs have measurements.

\subsection{Refactoring Operation Extraction}
\label{sec:miner}

To identify and quantify the transformations performed by the agent (\RQ{2}), we extract applied refactoring operations with RefactoringMiner (version 3.1.2)~\citep{Tsantalis:ICSE:2018:RefactoringMiner, Tsantalis:TSE:2020:RefactoringMiner2.0, Alikhanifard:TOSEM:2024:RefactoringMiner3.0}, which detects refactoring activity from the Abstract Syntax Tree (AST) differences between the pre- and post-refactoring versions. RefactoringMiner supports Java and Python, so this analysis covers those two subsets and not C++. It detects 103 distinct operation types.

We classify refactoring operations as High-, Medium-, or Low-level using the framework\footnote{\url{https://github.com/Mont9165/Agent_Refactoring_Analysis/blob/main/Refactoring_Level_Classification.md}} of \citet{horikawa2025agenticrefactoringempiricalstudy}, which groups operations by the scope of the change: Low-level operations are confined to a code block (e.g., Rename Variable, Extract Variable), High-level operations change signatures of classes, methods, or fields (e.g., Rename Method, Change Access Modifier, Move Method), and Medium-level operations change both a signature and the code block (e.g., Extract Method, Extract Class). The framework does not rank levels by maintainability impact. However, in our data, the High level is dominated by renames and access-modifier changes (Table~\ref{tab:ref_ops}). Therefore, when we describe operations as structural, we refer mostly to the Medium-level operations (extract method) that occur in our experiments. Appendix~\ref{sec:operation-levels} lists the level of every operation type observed in our data.

\subsection{Failure Taxonomy}
\label{sec:failure-taxonomy}

To analyze failures (\RQ{4}), we categorize failed refactorings by the first stage they fail at, into Compile Errors, Runtime Errors, Test Failures, and Exceptions (agent or tooling malfunctions, including runs that produce no refactored file). No separate compilation step was used for Python. Consequently, for cross-language comparability we classify the Python error classes that a compiler rejects in Java and C++ as Compile Errors (\texttt{SyntaxError}, \texttt{IndentationError}, \texttt{TabError}, \texttt{NameError}, \texttt{UnboundLocalError}, \texttt{ImportError}, and \texttt{ModuleNotFoundError}). Compile and runtime error messages are further mapped to causes with rule-based regular expressions. Each stage has a 20-minute timeout that are classified according to the stage in which they occur (runtime or compile time). The full mapping is in the replication package~\cite{replication}.

\subsection{Statistical Analysis}
\label{sec:stats}

We use exact McNemar tests for paired nominal outcomes (pass rate, improved CH rate, effective refactoring rate, per-smell remediation). For these tests we report Cohen's $g$ as effect size ($>0.05$ small, $>0.15$ medium, $>0.25$ large)~\citep{Cohen1992} alongside a 95\% confidence interval computed from the discordant pairs. Because we run one McNemar test per code smell, we control the family-wise error rate with a Holm correction within each language. For continuous outcomes (CH scores, turns, tokens), we use Wilcoxon signed-rank tests, reporting the effect size $r$ ($>0.1$ small, $>0.3$ moderate, $>0.5$ large)~\citep{Cohen1992}.

To evaluate the shift in refactoring levels (\RQ{2}), individual operations are not treated as independent observations, since a single task can emit multiple refactoring operations. We therefore aggregate to the task level and compare the paired shares with a Wilcoxon signed-rank test. The mean difference in per-task shares is reported with a 95\% bootstrap confidence interval ($2{,}000$ task-level resamples). We also include Cram\'er's $V$ as a descriptive summary of the pooled (High-, Medium-, and Low-level) operations.

To evaluate how initial code quality moderates guidance (\RQ{3}), we model the probability of an effective refactoring, pass rate and improvement in CH using Generalized Estimating Equations (GEE) logistic regressions. We cluster by task to account for the repeated, paired Basic and Guided runs. The model includes the prompt condition, initial CH, and their interaction. From this interaction, we also calculate the model-implied break-even point.

When evaluating failure-mode shifts (\RQ{4}), we account for the overlapping sets of failed refactorings across conditions by using a within-task permutation test ($10{,}000$ permutations) that randomly permutes the Basic and Guided outcomes within each task pair. The test is applied to each failure category's share among failed refactorings. The test-failure share is the pre-specified primary contrast, and the remaining categories are Holm-corrected within language.

\subsection{Visualization Design}
\label{sec:visual}
In relation to \RQ{1} and \RQ{3}, we present a visualization of refactoring outcomes that complements the statistical analysis of the aggregated results (Figure~\ref{fig:movement}). The tables establish the average effect of guidance, but they suppress the distribution behind it, e.g., how many files moved, how far, in which direction, and from what starting point. Since CH is a central construct in the investigation, we design the visualization to show per-file CH movement, ordered by initial CH.

As recommended by~\citet{munzner2009nested}, we state the assumptions upstream of our focus on the visual encoding. First, the data abstraction is, per task, (i) two quantitative CH values, the \textit{initial} and the \textit{post-refactoring} score, and (ii) a categorical CH outcome for the run (improved, reached the maximum, degraded, unchanged, or failed). Second, the visualization is designed to support (i) judging the direction and magnitude of movement for refactoring tasks, (ii) comparing the distribution of movement between Basic and Guided, and (iii) relating both to the initial CH.

The visual encoding maps each individual refactoring task to one horizontal segment spanning from its initial to its final CH. Magnitude and direction are thus read as length and orientation along a common scale. Rows are ordered by initial CH, which promotes the moderating variable of \RQ{3} to a visible vertical axis. 

Allocating one row per task places the design in the pixel-oriented family~\citep{keim2000pixel}, which trades the ability to identify any individual task for the ability to show all of them at once. At the page size of Figure~\ref{fig:movement}, this means we have to cluster neighboring refactoring tasks, which we will explain when presenting the resulting visualizations. 

Two outcome categories are visualized in the right margin. First, refactoring tasks that lead to failures (failed), e.g., compilation errors, run-time errors, or tests not passing, are presented as black bars, representing the fraction of failures among neighboring tasks. Second, refactoring tasks that pass the tests without moving the CH (unchanged) are presented analogously, as a light gray stacked bar. For these bars in the margin, a group of neighboring tasks covers about 140 tasks for C++ and Java, and 70 for Python.

We justify the chosen encoding from theoretical principles of information visualization rather than from empirical evidence, which \citet{merino2018slr} identify as the two strategies available for establishing the effectiveness of a software visualization. We did not conduct a user study, so the usefulness of the design is argued rather than validated. Finally, we stress that the visualizations are intended to support qualitative visual analysis of distributional patterns and individual movements (given adequate resolution) -- exact numerical results are instead reported in the tables.

\subsection{Conversation Logs}
\label{sec:logs-method}

As an exploratory qualitative complement, we examined a small sample of agent conversation logs. We randomly sampled seven sessions, four from Java (one Basic and three Guided) and three from Python (one Basic, two Guided). We noted recurring patterns in how the agent used tools, edited code, and decided when to terminate. No coding scheme was applied. Consequently, the observations are illustrative insights, and do not establish how representative or frequent these patterns are.

\section{Results}
\label{sec:results}

This section reports the results for each RQ, each ending with a boxed summary, followed by observations from sampled conversation logs.

\subsection{\RQ{1}: Impact of Metric Guidance on Refactoring Outcomes}
\label{sec:rq1}

Table~\ref{tab:functional_results} presents the refactoring outcomes. For functional correctness, the Basic runs yield higher pass rates than the Guided runs across all languages. Paired McNemar tests indicate that this is statistically significant ($p < 0.01$), but the effect sizes range from negligible to small ($0.03 \le g \le 0.13$), with the largest drop in Java, that decreased by $8.6$ pp.

\begin{table}[htbp]
    \centering
    \caption{Pass, improved-CH, and effective refactoring rates. The paired rate difference $\Delta$ (Guided $-$ Basic) is reported with a 95\% confidence interval from the discordant pairs.}
    \label{tab:functional_results}
    \begin{tabular}{@{} l c c c c c @{}}
        \toprule
        \textbf{Language} & \textbf{Basic (\%)} & \textbf{Guided (\%)} & \textbf{$\Delta$ (pp)} & \textbf{95\% CI} & \textbf{Cohen's $g$}\\
        \midrule
        \multicolumn{6}{@{}l}{\textbf{Pass Rate}} \\
        \midrule
        C++    & 62.1 & 59.8 & $-2.3$** & $[-3.6, -1.0]$ & 0.03 \\
        Java   & 77.2 & 68.6 & $-8.6$** & $[-9.7, -7.5]$ & 0.13 \\
        Python & 82.4 & 78.7 & $-3.7$** & $[-5.0, -2.3]$ & 0.08\\
        \midrule
        \multicolumn{6}{@{}l}{\textbf{Improved CH Rate}} \\
        \midrule
        C++    & 30.6 & 69.6 & $+39.0$** & $[+37.9, +40.2]$ & 0.40\\
        Java   & 46.9 & 81.8 & $+35.0$** & $[+33.9, +36.0]$ & 0.42\\
        Python & 43.0 & 77.2 & $+34.2$** & $[+32.7, +35.8]$ & 0.39\\
        \midrule
        \multicolumn{6}{@{}l}{\textbf{Effective Refactoring Rate}} \\
        \midrule
        C++    & 17.6 & 42.8 & $+25.2$** & $[+24.1, +26.4]$ & 0.32\\
        Java   & 33.6 & 54.8 & $+21.1$** & $[+19.9, +22.4]$ & 0.24\\
        Python & 34.4 & 60.3 & $+25.9$** & $[+24.2, +27.6]$ & 0.30\\
        \bottomrule
        \multicolumn{6}{@{}l}{\footnotesize Significance: * $p < 0.05$, ** $p < 0.01$.} \\
    \end{tabular}
\end{table}

The CH component moves in the opposite direction. Guidance raises the share of runs that strictly improve CH from $30.6$\% to $69.6$\% in C++, from $46.9$\% to $81.8$\% in Java, and from $43.0$\% to $77.2$\% in Python ($+34$--$39$ pp, $p < 0.01$), with large effect sizes ($0.39 \le g \le 0.42$).

The condition for an effective refactoring requires both a passing test suite (a proxy for behavior preservation) and an improved CH (a proxy for code quality). Guidance lowers the pass rate by $2.3$--$8.6$ pp but raises the improved-CH rate by $34.2$--$39.0$ pp. The effective refactoring rate increases by $+25.2$ pp in C++, $+21.1$ pp in Java, and $+25.9$ pp in Python ($p < 0.01$), with medium to large effect sizes ($0.24 \le g \le 0.32$). Consequently, the Guided agent trades a small loss in correctness for a larger gain in impactful refactorings that improve internal code quality.

\begin{table}[htbp]
    \centering
    \caption{CH improvements from initial to refactored (mean $\pm$ SD and Wilcoxon $r$) over the tasks with a CH score for the refactored file in both runs ($9{,}207$ C++, $9{,}700$ Java, $4{,}967$ Python). The final columns report paired within-task CH differences (mean $\Delta$, 95\% CI, and Wilcoxon $r$).}
    \label{tab:quality_results}
    \footnotesize
    \setlength{\tabcolsep}{4pt}
    \begin{tabular}{@{} l c c c c c c c @{}}
        \toprule
        & & \multicolumn{2}{c}{\textbf{Basic}} & \multicolumn{2}{c}{\textbf{Guided}} & \multicolumn{2}{c}{\textbf{Guided $-$ Basic}} \\
        \cmidrule(lr){3-4} \cmidrule(lr){5-6} \cmidrule(lr){7-8}
        \textbf{Language} & \textbf{Init CH} & \textbf{CH} & \textbf{$r$} & \textbf{CH} & \textbf{$r$} & \textbf{$\Delta$ [95\% CI]} & \textbf{$r$} \\
        \midrule
        \textbf{C++}      & $7.87 \pm 1.25$ & $7.99 \pm 1.18$ & 0.09** & $8.54 \pm 1.13$ & 0.68** & $+0.55~[0.53, 0.56]$** & 0.64 \\
        \textbf{Java}     & $8.08 \pm 1.08$ & $8.40 \pm 1.01$ & 0.55** & $8.95 \pm 0.87$ & 0.81** & $+0.55~[0.54, 0.57]$** & 0.68 \\
        \textbf{Python}   & $9.03 \pm 0.64$ & $9.15 \pm 0.67$ & 0.31** & $9.52 \pm 0.60$ & 0.75** & $+0.38~[0.36, 0.39]$** & 0.67 \\
        \bottomrule
        \multicolumn{8}{@{}l}{\footnotesize Significance: * $p < 0.05$, ** $p < 0.01$.} \\
    \end{tabular}
\end{table}

Table~\ref{tab:quality_results} shows the corresponding internal quality improvements. Both Basic and Guided runs improve CH significantly over the initial state ($p < 0.01$), but the magnitudes differ. In C++, the Basic agent's improvement is negligible ($r = 0.09$), while the Guided agent reaches a large effect ($r = 0.68$). In Python, the contrast is moderate versus large ($r = 0.31$ vs.\ $r = 0.75$). In Java, both effects are large, but guidance is stronger ($r = 0.81$ vs.\ $r = 0.55$). Compared directly, Guided runs end with a higher CH than Basic runs by about $+0.55$ in C++ and Java and $+0.38$ in Python.




\begin{figure}[htbp]
\centering
\includegraphics[width=\textwidth]{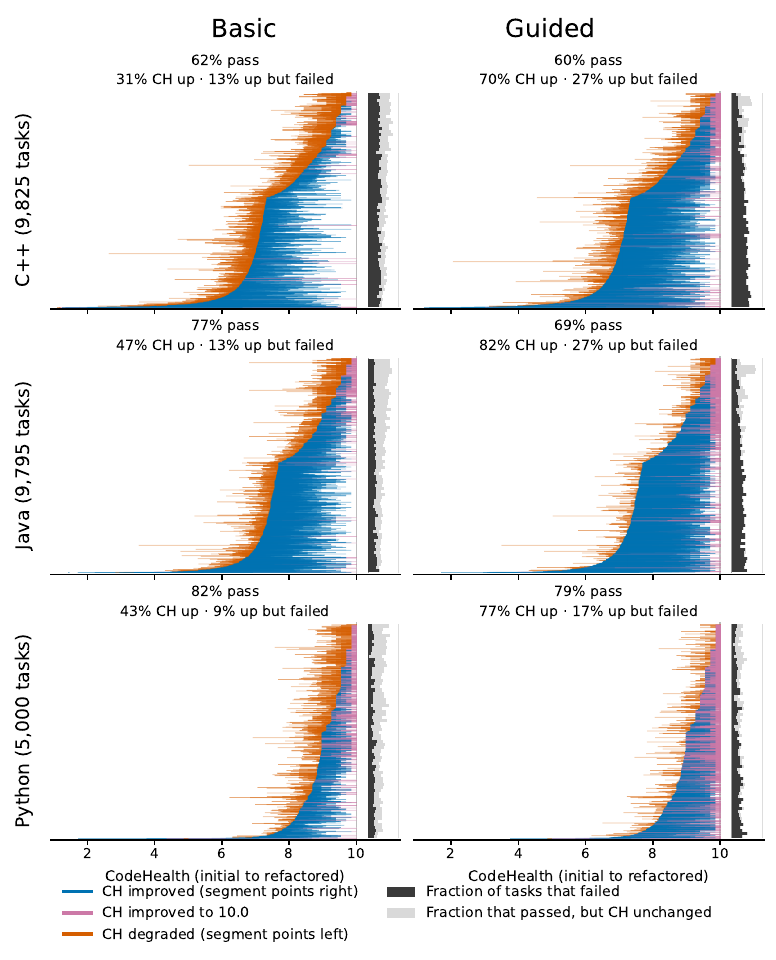}
\caption{CH trajectories for refactorings, Basic (left) and Guided (right). Horizontal segments span the initial to final CH score, sorted by initial CH (high to low, top to bottom). Colored by improvement (blue), degradation (orange), or reaching a perfect CH of 10 (pink). The stacked bars to the right of each visualization show, among tasks with similar initial CH, the fraction that failed (dark) and the fraction that passed without changing CH (light).}
\label{fig:movement}
\end{figure}

Figure~\ref{fig:movement} shows the CH movement of all $24{,}620$ paired tasks using the design presented in Section~\ref{sec:visual}. Rows are ordered by initial CH, and the order is identical in the Basic and Guided visualizations of a language, i.e., similar tasks appear at the same height in both and can be compared directly. Each visualization carries up to 9{,}825 rows in a few hundred device pixels, which means that each visible row therefore covers about 8--15 tasks with similar initial CH. Every task occupies a row, without sampling or averaging, but the overlap forms an envelope: the colored area reaches as far as the largest CH change among neighboring tasks. The figure thus conveys the direction and the reach of the movement across the CH range, whereas magnitudes per task and rates are reported in Tables~\ref{tab:functional_results} and~\ref{tab:quality_results}.

Visual inspection reveals that the aggregate improvement under guidance corresponds to a broad shift in the distribution of refactoring outcomes rather than to a small number of extreme cases. Across all three languages, Guided produces a visibly denser set of blue rightward movements, and also more tasks that reach the maximum CH of 10 (pink). The extent of the movements naturally depends on initial CH. Tasks starting at lower CH have more headroom, and can thus improve by several points. This overall pattern is visible for C++, Java, and Python. In Python, the number of pink trajectories reaching CH 10 is particularly pronounced. 

We see that the broader positive CH movement under Guided is accompanied by higher failure rates, as reported in Table~\ref{tab:functional_results}. For every language, the black bars in the right margin generally extend further under Guided than under Basic. One notable exception occurs among the healthiest C++ tasks, where Guided instead shows a lower fraction of failures. Across the languages, failures under Guided are concentrated more strongly among tasks starting at lower CH, visible as longer black bars toward the lower-CH portions of the panels. This visual pattern indicates that the difference between Basic and Guided is not uniform across initial CH and is consistent with the interaction examined in \RQ{3}.

\finding{1}{Metric guidance raises the effective refactoring rate, measured under the guiding metric, by $21$--$26$ pp across all three languages (medium-to-large effects), driven by a $34$--$39$ pp rise in the share of runs that improve CH, at the cost of a small reduction in pass rate ($2.3$--$8.6$ pp, negligible-to-small effects).}

\subsection{\RQ{2}: Changes in Refactoring Operations and Smell Remediation}
\label{sec:rq2}

Table~\ref{tab:ref_ops} summarizes the refactoring operations detected between the initial and refactored code for the Java and Python subsets (the version of RefactoringMiner used does not support C\texttt{++}), grouped by refactoring level~\citep{horikawa2025agenticrefactoringempiricalstudy}.

\begin{table}[htbp]
    \centering
    \caption{Share of refactoring operations (\%) per level and for the ten most frequent operations of each language. $\Delta$ is the pp difference (Guided $-$ Basic). The shares are over $172{,}485$ Basic and $77{,}152$ Guided operations in Java, and $70{,}826$ and $39{,}281$ in Python.}
    \label{tab:ref_ops}
    \footnotesize
    \begin{tabular}{@{} l rrr rrr @{}}
        \toprule
        & \multicolumn{3}{c}{\textbf{Java}} & \multicolumn{3}{c}{\textbf{Python}} \\
        \cmidrule(lr){2-4} \cmidrule(lr){5-7}
        \textbf{Operation} & \textbf{Basic} & \textbf{Guided} & \textbf{$\Delta$ (pp)} & \textbf{Basic} & \textbf{Guided} & \textbf{$\Delta$ (pp)} \\
        \midrule
        \textit{High-level}      & \textit{30.88} & \textit{19.56} & \textit{$-11.32$} & \textit{19.46} & \textit{19.15} & \textit{$-0.32$} \\
        Rename Method             & 3.02  & 1.48  & $-1.54$  & 9.95  & 5.62  & $-4.34$  \\
        Rename Attribute             & 5.65  & 1.12  & $-4.53$  & 3.94  & 1.08  & $-2.86$  \\
        Change Attribute Access Modifier & 6.88  & 2.34  & $-4.54$  & --    & --    & --       \\
        Parameterize Variable             & 4.07  & 9.30  & $+5.24$  & 1.23  & 5.11  & $+3.88$  \\
        Change Method Access Modifier & 4.03  & 1.60  & $-2.42$  & 0.17  & 0.29  & $+0.11$  \\
        Add Parameter               & 1.49  & 0.56  & $-0.94$  & 2.03  & 2.61  & $+0.58$  \\
        \midrule
        \textit{Medium-level}    & \textit{12.11} & \textit{48.67} & \textit{$+36.56$} & \textit{10.97} & \textit{31.66} & \textit{$+20.69$} \\
        Extract Method            & 6.44  & 45.20 & $+38.76$ & 5.07  & 24.07 & $+19.00$ \\
        Extract Attribute            & 2.75  & 0.48  & $-2.27$  & 0.78  & 0.12  & $-0.66$  \\
        Change Return Type        & 0.14  & 0.14  & $+0.01$  & 3.08  & 4.61  & $+1.53$  \\
        \midrule
        \textit{Low-level}       & \textit{57.02} & \textit{31.77} & \textit{$-25.24$} & \textit{69.57} & \textit{49.20} & \textit{$-20.37$} \\
        Rename Variable              & 41.04 & 16.07 & $-24.96$ & 56.84 & 35.97 & $-20.87$ \\
        Rename Parameter            & 8.79  & 2.25  & $-6.55$  & 6.23  & 2.70  & $-3.54$  \\
        Extract Variable             & 2.42  & 3.49  & $+1.07$  & 4.14  & 5.09  & $+0.95$  \\
        Inline Variable              & 0.51  & 1.33  & $+0.83$  & 0.98  & 2.04  & $+1.06$  \\
        \bottomrule
    \end{tabular}
\end{table}

The proportional distributions show a contrast in strategy. Without explicit guidance, the Basic agent favors Low-level refactorings ($57.02$\% in Java, $69.57$\% in Python), where \textit{Rename Variable} accounts for $41.04$\% of all Basic operations in Java and $56.84$\% in Python. High-level changes that do not alter core logic, such as \textit{modifying access modifiers} or \textit{renaming attributes}, form the secondary share ($30.88$\% in Java and $19.46$\% in Python) of refactoring operations. This mirrors the consistency-oriented profile reported for agents by~\citet{horikawa2025agenticrefactoringempiricalstudy} and  \citet{ottenhof2026agentsrefactorempiricalstudy}.

Under metric guidance, the agent's behavior shifts from superficial edits toward structural improvements. Medium-level operations rise from $12.11$\% to $48.67$\% of all refactorings in Java and from $10.97$\% to $31.66$\% in Python, driven primarily by the \textit{Extract Method} ($45.20$\% of Guided operations in Java, $24.07$\% in Python) operation. The share of Low-level operations drops to $31.77$\% in Java and $49.20$\% in Python.

\begin{table}[htbp]
    \centering
    \caption{Task-level shift in the mean per-task share of operations at each refactoring level, over all detected operations for all refactorings with operations in both runs ($9{,}023$ Java- and $4{,}284$ Python refactorings). $\Delta$ is the paired difference with a 95\% CI, and $r$ the effect size.}
    \label{tab:level_shift}
    \begin{tabular}{@{} l l c c c c c @{}}
        \toprule
        \textbf{Language} & \textbf{Level} & \textbf{Basic (\%)} & \textbf{Guided (\%)} & \textbf{$\Delta$ (pp)} & \textbf{95\% CI} & \textbf{$r$} \\
        \midrule
        \multirow{3}{*}{\textbf{Java}}
          & High   & $23.6$ & $13.2$ & $-10.4$** & $[-11.0, -9.8]$  & 0.39 \\
          & Medium & $15.9$ & $58.3$ & $+42.3$** & $[+41.6, +43.0]$ & 0.80 \\
          & Low    & $60.4$ & $28.5$ & $-31.9$** & $[-32.7, -31.1]$ & 0.68 \\
        \midrule
        \multirow{3}{*}{\textbf{Python}}
          & High   & $20.4$ & $16.0$ & $-4.4$**  & $[-5.3, -3.6]$   & 0.19 \\
          & Medium & $15.2$ & $41.9$ & $+26.7$** & $[+25.7, +27.8]$ & 0.67 \\
          & Low    & $64.4$ & $42.1$ & $-22.3$** & $[-23.5, -21.1]$ & 0.52 \\
        \bottomrule
        \multicolumn{7}{@{}l}{\footnotesize Significance: * $p < 0.05$, ** $p < 0.01$.} \\
    \end{tabular}
\end{table}

The choice of operation depends on the solution that is refactored. Therefore, we compare each solutions share of operation at each level across experiments (Table~\ref{tab:level_shift}). In this setting, guidance raises the Medium-level share by $\sim$$4\times$ in Java and by $\sim$$3\times$ in Python ($r = 0.80$ and $0.67$, $p < 0.01$). For Java, this contrast comes from Low-level ($-31.9$ pp) and High-level ($-10.4$ pp) operations. For Python, decrease in Low-level ($-22.3$ pp) dominates. Within the High-level category, renames decreased in both languages and access-modifier changes in Java, while \textit{Parameterize Variable}, a byproduct of method extraction, increases (Table~\ref{tab:ref_ops}). Over the pooled operation count instead, the association between prompt condition and level category gives a Cram\'er's $V$ of $0.399$ for Java and $0.262$ for Python.

\begin{figure}[htbp]
    \centering
    \includegraphics[width=\textwidth]{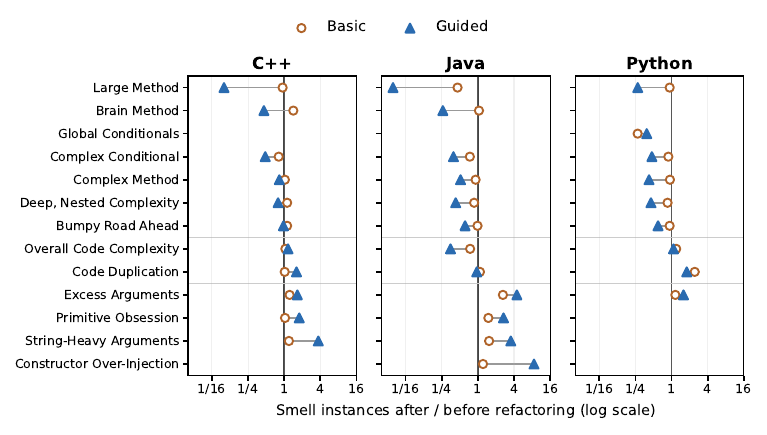}
    \caption{Smell instances after refactoring relative to before, Basic (circle) and Guided (triangle), over the solutions with a CH score of both refactored files. A ratio below 1 means net remediation, above 1 net introduction (net reduction $= 100 \times (1 - \text{ratio})$\%). Smells with fewer than 20 instances before refactoring in a language are omitted.}
    \label{fig:smells}
\end{figure}

Next we examine how this operational shift translates into smell remediation. Figure~\ref{fig:smells} reports the code smell instances before and after a refactoring, and Table~\ref{tab:mcnemar_tests} tests, per smell, if one condition remediates it more than the other. The guided agent achieves large net reductions of complexity-related code smells, with reduction of $72.7$--$96.1$\% for \textit{Large Method} and  $51.6$--$60.8$\% for \textit{Complex Conditional}. Across all languages, the per-smell McNemar tests confirm the shift. The Guided run remediates more code smells with large effect sizes ($0.42 \le g \le 0.48$, Holm-corrected $p < 0.01$). Therefore, the shift toward Medium-level operations carries over to the granular smell level.

\begin{table}[htbp]
    \centering
    \caption{McNemar's test for smell remediation, over the tasks with a valid CH report of both refactored files. Basic and Guided are the discordant pairs, with Cohen's $g$ as effect size ($>0.05$ small, $>0.15$ medium, $>0.25$ large).}
    \label{tab:mcnemar_tests}
    \footnotesize
    \setlength{\tabcolsep}{4pt}
    \begin{tabular}{@{} l rrc rrc rrc @{}}
        \toprule
        & \multicolumn{3}{c}{\textbf{C++}} & \multicolumn{3}{c}{\textbf{Java}} & \multicolumn{3}{c}{\textbf{Python}} \\
        \cmidrule(lr){2-4} \cmidrule(lr){5-7} \cmidrule(lr){8-10}
        \textbf{Target Smell} & \textbf{Basic} & \textbf{Guided} & \textbf{$g$} & \textbf{Basic} & \textbf{Guided} & \textbf{$g$} & \textbf{Basic} & \textbf{Guided} & \textbf{$g$} \\
        \midrule
        Large Method            & 33  & $1{,}974$** & 0.48 & 24  & $712$**     & 0.47 & 6   & $69$**      & 0.42 \\
        Complex Conditional     & 124 & $1{,}515$** & 0.42 & 136 & $1{,}584$** & 0.42 & 20  & $311$**     & 0.44 \\
        Complex Method          & 156 & $3{,}129$** & 0.45 & 196 & $3{,}870$** & 0.45 & 75  & $1{,}513$** & 0.45 \\
        Deep, Nested Complexity & 81  & $1{,}915$** & 0.46 & 104 & $2{,}334$** & 0.46 & 71  & $911$**     & 0.43 \\
        Bumpy Road Ahead        & 144 & $2{,}539$** & 0.45 & 234 & $3{,}291$** & 0.43 & 136 & $1{,}594$** & 0.42 \\
        Code Duplication        & 83  & $256$**     & 0.26 & 148 & $955$**     & 0.37 & 19  & $49$**      & 0.22 \\
        Primitive Obsession     & 184 & $89$**      & 0.17 & 230 & $31$**      & 0.38 & --  & --          & --   \\
        \bottomrule
        \multicolumn{10}{@{}l}{\footnotesize Significance: * $p < 0.05$, ** $p < 0.01$, Holm-corrected within language.} \\
    \end{tabular}
\end{table}

However, structural decomposition introduces language-specific trade-offs. As the agent extracts methods to reduce complexity, it risks introducing smells related to parameter passing and coupling. Figure~\ref{fig:smells} shows a net introduction of \textit{Excess Arguments} in Java ($-346.8$\%, growing from $1{,}080$ to $4{,}825$ instances) compared to C++ ($-65.8$\%) and Python ($-57.9$\%), alongside \textit{String-Heavy Arguments} ($-253.8$\%) and \textit{Constructor Over-Injection} ($-760.9$\%) in Java. The latter two are driven by small absolute bases and should be read as directional signals. \textit{Primitive Obsession} is the only tested smell where the Basic agent outperforms the Guided agent, in both C++ ($184$ vs.\ $89$) and Java ($230$ vs.\ $31$). While the agent successfully breaks down large functions, it struggles to create clean data transfer between the newly extracted components.

The same patterns occur for the Basic experiments (Figure~\ref{fig:smells}). Remediation of complexity smells is more infrequent (\textit{Large Method} $+5.8$\% in C++ and $+54.2$\% in Java versus $+90.0$\% and $+96.1$\% under Guided runs) and sometimes negative (\textit{Deep, Nested Complexity} $-12.1$\% in C++). The coupling smells, in turn, are already introduced without guidance, at $\sim$$0.35$--$0.5\times$ the Guided rate in C++ and Java (\textit{Excess Arguments} $-23.2$\% and $-161.4$\% versus $-65.8$\% and $-346.8$\%). Guidance then amplifies an existing tendency of the agent's decompositions rather than creating it.

\finding{2}{Metric guidance shifts refactorings away from low-level edits, which fall from $57$\% to $32$\% of operations in Java and from $70$\% to $49$\% in Python, toward structural, \textit{Extract Method}-dominated transformations, which rise from $12$\% to $49$\% in Java and from $11$\% to $32$\% in Python. Remediation of complexity smells rises across all languages, but the decomposition introduces new coupling smells (excess and string-heavy arguments).}

\subsection{\RQ{3}: Moderating Effect of Initial CH}
\label{sec:rq3}

Metric guidance improves outcomes on average, but its success depends on the initial quality of the file to refactor. To formalize this moderation, we fit GEE logistic regressions of the effective-refactoring, pass rate and CH improvement probability on the prompt condition, initial CH, and their interaction, clustered by solution. The model specifications (Equation~\ref{eq:gee}) and the full coefficient estimates (Table~\ref{tab:gee_regression}) are given in Appendix~\ref{sec:gee-details}. Figure~\ref{fig:rq3} shows this across languages, plotting the GEE modeled probability of an effective refactoring (top row) together with its two components, strictly increasing CH (middle row) and passing the test suite (bottom row), against initial CH.

\begin{figure}[htbp]
    \centering
    \includegraphics[width=\textwidth]{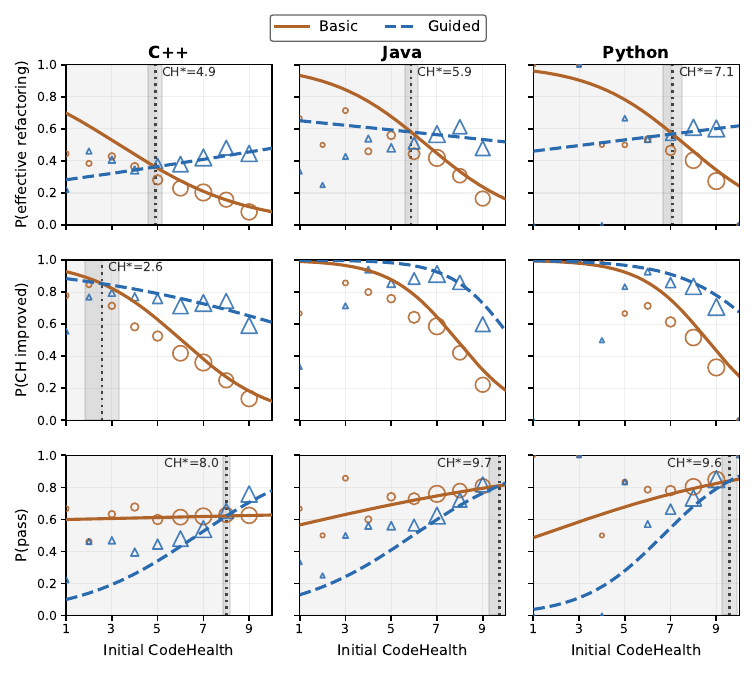}
    \caption{Effective refactoring (top row) and its two components, increasing CH (middle row) and passing the tests (bottom row), by initial CH. Curves show GEE predictions and markers denote observed bin rates, sized by task count (triangle for Guided and circle for Basic). The dotted line ($\text{CH}^*$, with 95\% CI band) and left shading indicate the region where the model expects guidance to be worse than Basic. The unconstrained curves may cross these bounds where they extrapolate into the sparse low-CH region.}
    \label{fig:rq3}
\end{figure}

For effective refactoring rate, the two experimental conditions respond to initial CH in opposite directions. Under the Basic instruction, the probability of an effective refactoring declines as initial CH increases (in C++, from $0.70$ at CH 1 to $0.11$ at CH 9), as healthier files inherently offer less headroom for structural improvement. Conversely, metric guidance stabilizes this probability across CH. Consequently, the performance trajectories intersect at language-specific break-even points.

The interaction term is positive and significant ($p < 0.01$) across all configurations ($+0.455$ in C++, $+0.415$ in Java, $+0.554$ in Python, $+0.338$ overall). This shows that the relative advantage of metric guidance scales with initial CH. From these coefficients, we derive the model-implied break-even points shown in Figure~\ref{fig:rq3}, ranging from $4.9$ in C++ to $7.1$ in Python.

Below these thresholds, a Guided run is less likely to generate a effective refactoring than a Basic run. However, guidance still improves CH, but pushing the agent toward structural transformation risks altering behaviour, particularly in degraded code~\cite{borg2026codemachinesjusthumans}. The Basic agent keeps to shallow edits, that reduces the risk of altering behaviour. Consequntly, this is more likely to produce an effective refactoring in degraded code, even though the magnitude of CH improvement is smaller. This suggests that it may be more difficult to perform structural transformations in low code quality for the agents, where intervention might be more suitable before AI uplifting.


The model-implied impact of guidance ranges from $-14.0$ pp in Python at CH 6 to $+34.1$ pp in C++ at CH 9, and each language turns positive above its own break-even point. Below CH 5, the estimates rest on few solutions across languages. In Python, the effect at CH 5 ($-25.0$ pp) is extrapolated from $n=6$ solutions and is therefore only read as indicative.

To ensure this moderation effect is a behavioral shift rather than a statistical artifact, we validated the interaction in two ways. First, we decomposed the composite ``effective refactoring'' criterion by evaluating the model against the pass rate alone (Figure~\ref{fig:rq3}, bottom row). The positive interaction remains significant ($+0.371$ in C++, $+0.249$ in Java, $+0.370$ in Python; all $p < 0.01$), showing that the trend is not only a byproduct of our success definition, under which high-CH files have little headroom. The middle row of Figure~\ref{fig:rq3} reveals that metric guidance raises the probability of improving the CH score across almost the entire CH quality range, with the curves crossing at CH $2.6$ in C++ and never in Java or Python. Therefore, the break-even thresholds are driven mostly by lower pass rate. On degraded code, the Guided agent still improves the metric but breaks behaviour more often. Furthermore, among behaviour preserving refactorings below the break-even point, the median CH gain is $+3.29$ for Guided runs compared to $+1.25$ for Basic in C++ and $+2.85$ against $+0.94$ in Java, and $+2.13$ against $+1.15$ in Python. Thus, what degrades below the break-even point is the likelihood for preserving behavior~\cite{borg2026codemachinesjusthumans} and not the value of the transformation performed.

Second, we ruled out file length as a potential confounder. SLoC remains relatively flat across the CH distribution (Appendix~\ref{sec:sloc}), and is nearly independent of initial CH (Spearman $|\rho| \le 0.11$). Apart from the MCP server failures, which are more frequent on larger files (median $181$ vs.\ $127$ lines in C++), we attribute the Guided regression on degraded code on structural entanglement rather than size. Entangled code does, however, lead to more token usage during sessions (median $221$k vs.\ $206$k tokens below and above CH 7 in C++), and the resulting context pressure~\citep{liu-etal-2024-lost} may contribute to the effect.


Together, these results extend the AI-Friendliness hypothesis~\citep{borg2026codemachinesjusthumans, thoughtworks2025} to agentic settings. Metric guidance amplifies an agent's capability on comprehensible code, but forcing agents to structurally decompose severely degraded code increases the risk of behavioral regressions.

\finding{3}{The benefit of metric guidance is significantly moderated by initial CH (GEE interaction $p < 0.01$ in all languages). Model-implied break-even points range from CH $4.9$ $[4.6, 5.2]$ in C++, $5.9$ $[5.6, 6.2]$ in Java, and $7.1$ $[6.7, 7.5]$ in Python. Below them, Guided runs produce fewer effective refactorings, as structural transformations more often break behavior, but successful transformations still deliver larger maintainability improvements than the shallower edits of the Basic agent.
}

\subsection{\RQ{4}: Shifts in Failure Modes}
\label{sec:rq4}

\begin{figure}[htbp]
    \centering
    \includegraphics[width=\textwidth]{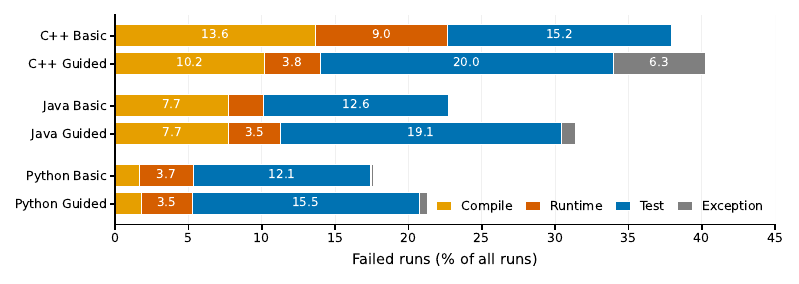}
    \caption{Failed refactoring runs as a share of all runs, by failure category. The remaining runs pass.}
    \label{fig:failure_modes}
\end{figure}

Metric guidance alters the distribution of failure modes across all languages (Figure~\ref{fig:failure_modes}). The shift shared by all languages is the rise in Test failures, which grow from $15.2$\% to $20.0$\% of all refactorings in C++, $12.6$\% to $19.1$\% in Java, and $12.1$\% to $15.5$\% in Python. However, because Guided runs have a lower overall pass rate, comparing absolute percentages obscures whether the agent's actual coding behavior changed, or if it generated more broken code overall.

To isolate the behavioral shift, Table~\ref{tab:failure_comp} evaluates the relative composition of failures among unsuccessful refactorings. Within this failure profile, a shift from syntactic and execution errors toward semantic logic errors emerges. The relative share of Test failures grows significantly across all three languages ($+9.5$ pp in C++, $+5.6$ pp in Java, $+3.9$ pp in Python). Conversely, the crash share falls in C++ (Runtime $-14.4$ pp) and Python (Runtime $-4.8$ pp), while C++ and Java shed early-stage errors (Compile $-10.6$ and $-9.2$ pp). Notably, Java's Runtime share among failed refactorings remains statistically unchanged ($+0.7$ pp).

The Exception category also rises under guidance (C++ $+15.6$ pp, Java $+2.9$ pp, Python $+1.7$ pp). In C++ and Java, most Guided exceptions ($575$ of $618$ and $92$ of $93$) are harness failures related to the MCP server, a non-zero exit with the MCP subprocess in the trace, before any refactoring completed. The remaining C++ exceptions and most Python exceptions ($20$ of $27$) are agent timeouts before a refactoring was produced. In Python, these occur in the longest sessions (median $14.5$ vs.\ $10$ interaction turns), consistent with the overhead reported in \RQ{5}. Therefore, the Exception shift is an effect of limitations in infrastructure and session-length.

\begin{table}[htbp]
    \centering
    \caption{Composition of failed refactoring attempts by failure category. $\Delta$ is the Guided $-$ Basic shift, tested with a within-task permutation test ($10{,}000$ permutations).}
    \label{tab:failure_comp}
    \begin{tabular}{@{} l l rr rr r @{}}
        \toprule
        & & \multicolumn{2}{c}{\textbf{Basic}} & \multicolumn{2}{c}{\textbf{Guided}} & \\
        \cmidrule(lr){3-4} \cmidrule(lr){5-6}
        \textbf{Language} & \textbf{Category} & \textbf{$N$} & \textbf{\%} & \textbf{$N$} & \textbf{\%} & \textbf{$\Delta$ (pp)} \\
        \midrule
        \multirow{4}{*}{\textbf{C++}}
          & Compile   & $1{,}341$ & $36.0$ & $1{,}002$ & $25.3$ & $-10.6$**  \\
          & Runtime   & $887$     & $23.8$ & $370$     & $9.4$  & $-14.4$**  \\
          & Test      & $1{,}498$ & $40.2$ & $1{,}964$ & $49.7$ & $+9.5$**   \\
          & Exception & $0$       & $0.0$  & $618$     & $15.6$ & $+15.6$**  \\
        \midrule
        \multirow{4}{*}{\textbf{Java}}
          & Compile   & $756$     & $33.9$ & $759$     & $24.7$ & $-9.2$**   \\
          & Runtime   & $236$     & $10.6$ & $347$     & $11.3$ & $+0.7$\phantom{**}     \\
          & Test      & $1{,}237$ & $55.4$ & $1{,}874$ & $61.0$ & $+5.6$**   \\
          & Exception & $3$       & $0.1$  & $93$      & $3.0$  & $+2.9$**   \\
        \midrule
        \multirow{4}{*}{\textbf{Python}}
          & Compile   & $81$      & $9.2$  & $89$      & $8.4$  & $-0.8$\phantom{**}     \\
          & Runtime   & $185$     & $21.0$ & $173$     & $16.3$ & $-4.8$**   \\
          & Test      & $606$     & $68.9$ & $774$     & $72.8$ & $+3.9$*    \\
          & Exception & $7$       & $0.8$  & $27$      & $2.5$  & $+1.7$**   \\
        \bottomrule
        \multicolumn{7}{@{}l}{\footnotesize Significance: * $p < 0.05$, ** $p < 0.01$.} \\
    \end{tabular}
\end{table}

A potential confounder for this shift is edit volume (churn), as generating more code increases the surface area for logic errors. Churn shows that Guided runs change fewer lines than Basic runs in every language (median $179$ vs.\ $226$ changed lines in C++, $103$ vs.\ $139$ in Java, and $88$ vs.\ $112$ in Python), with moderate-to-large paired effects ($r = 0.37$--$0.58$, $p < 0.01$). Basic runs perform $\sim$$1.8$--$2.2\times$ more refactorings, dominated by renames that occur across whole solutions. Guided runs are more concentrated, transforming narrower areas pointed out to contain design flaws by the CH tool report. The rise in failures also reflects the change in operations of the guided edits, structural transformations in place of the shallow formatting and mass renaming typical of the Basic prompt (Section~\ref{sec:rq2}) that inherently increases the risk.

Granular error categorization (Table~\ref{tab:error_causes}) reveals the underlying mechanics of this shift. In C++, \textit{Missing Symbol/Identifier} errors make up $62.7$\% of Guided compile errors, up from $45.0$\% under Basic, indicating that the agent successfully generates new extracted components but sometimes fails to properly integrate them into the compilation unit.

\begin{table}[htbp]
    \centering
    \caption{Top 3 causes of compile and runtime errors by language and prompt, as a share (\%) of all compile or runtime errors of that run set.}
    \label{tab:error_causes}
    \footnotesize
    \setlength{\tabcolsep}{4pt}
    \begin{tabular}{@{} l l l r l r @{}}
        \toprule
        \textbf{Language} & \textbf{Prompt} & \textbf{Compile error} & \textbf{\%} & \textbf{Runtime error} & \textbf{\%} \\
        \midrule
        \multirow{6}{*}{\textbf{C++}}
            & \multirow{3}{*}{Basic}  & Missing Symbol/Identifier & 45.0 & Segmentation Fault      & 84.9 \\
            &                         & Ambiguous Reference       & 10.2 & Timeout                 & 10.4 \\
            &                         & Signature Mismatch        & 7.2  & Memory Corruption/Alloc & 2.3  \\
            \cmidrule{2-6}
            & \multirow{3}{*}{Guided} & Missing Symbol/Identifier & 62.7 & Segmentation Fault      & 60.5 \\
            &                         & Signature Mismatch        & 8.0  & Timeout                 & 26.8 \\
            &                         & Type Mismatch             & 6.2  & Memory Corruption/Alloc & 3.8  \\
        \midrule
        \multirow{6}{*}{\textbf{Java}}
            & \multirow{3}{*}{Basic}  & Missing Symbol/Identifier & 66.7 & Index Out of Bounds     & 49.2 \\
            &                         & Type Mismatch             & 9.9  & Missing/Null Reference  & 16.5 \\
            &                         & Signature Mismatch        & 4.6  & Input Parsing           & 12.3 \\
            \cmidrule{2-6}
            & \multirow{3}{*}{Guided} & Missing Symbol/Identifier & 43.6 & Index Out of Bounds     & 42.7 \\
            &                         & Type Mismatch             & 27.5 & Missing/Null Reference  & 19.0 \\
            &                         & Exception Handling        & 11.9 & Timeout                 & 17.0 \\
        \midrule
        \multirow{6}{*}{\textbf{Python}}
            & \multirow{3}{*}{Basic}  & Name Binding              & 88.9 & Type / Value Error      & 34.1 \\
            &                         & Syntax/Indentation        & 6.2  & Index Out of Bounds     & 24.3 \\
            &                         & Import Resolution         & 4.9  & Missing/Null Reference  & 20.5 \\
            \cmidrule{2-6}
            & \multirow{3}{*}{Guided} & Name Binding              & 92.1 & Type / Value Error      & 29.5 \\
            &                         & Syntax/Indentation        & 7.9  & Index Out of Bounds     & 24.3 \\
            &                         &                           &      & Timeout                 & 20.8 \\
        \bottomrule
    \end{tabular}
\end{table}

Furthermore, runtime errors shift away from primitive memory violations, as C++ \textit{Segmentation Faults} drop from $84.9$\% to $60.5$\% of runtime errors, displaced by \textit{Timeouts}, which also enter the top three runtime errors for Java ($17.0$\%) and Python ($20.8$\%).

Together, these patterns suggest that metric guidance directs the agent away from cosmetic edits that cause immediate syntactic crashes, while the structural transformations it encourages instead risk semantic logic errors and non-terminating loops, as control flow and parameter passing are reorganized.

\finding{4}{Under metric guidance, the failure profile shifts toward semantic errors. The Test-failure share rises in all languages, the crash share falls in C++ and Python and the compile-error share also decreases in C++ and Java, and C++ segmentation faults drop from $85$\% to $61$\% of runtime errors as timeouts grow. This is consistent with how Guidance encourages structural changes that increases risks for semantic faults rather than syntactic.}

\subsection{\RQ{5}: Computational Overhead}
\label{sec:rq5}

\begin{table}[htbp]
    \centering
    \caption{Token usage and interaction overhead per task (mean $\pm$ SD) over the tasks with available session statistics in both runs, with Wilcoxon signed-rank effect size $r$ for the paired Guided-vs.-Basic.}
    \label{tab:tokens}
\begin{tabular}{ l c c c c }
    \toprule
    \textbf{Language} & \textbf{Basic} & \textbf{Guided} & \textbf{Factor} & \textbf{$r$} \\
    \midrule
    \multicolumn{5}{l}{\textbf{Total Tokens}} \\
    \midrule
    C++    & $40{,}402 \pm 16{,}692$ & $213{,}673 \pm 62{,}315$ & $5.3\times$** & 0.87 \\
    Java   & $36{,}808 \pm 15{,}403$ & $197{,}518 \pm 77{,}121$ & $5.4\times$** & 0.87 \\
    Python & $54{,}493 \pm 19{,}021$ & $161{,}660 \pm 67{,}686$ & $3.0\times$** & 0.86 \\
    \midrule
    \multicolumn{5}{l}{\textbf{Output Tokens}} \\
    \midrule
    C++    & $2{,}770 \pm 1{,}275$ & $5{,}068 \pm 1{,}777$ & $1.8\times$** & 0.78 \\
    Java   & $2{,}520 \pm 1{,}192$ & $4{,}333 \pm 1{,}864$ & $1.7\times$** & 0.71 \\
    Python & $1{,}760 \pm 564$     & $3{,}435 \pm 1{,}559$ & $2.0\times$** & 0.79 \\
    \midrule
    \multicolumn{5}{l}{\textbf{Interaction Turns}} \\
    \midrule
    C++    & $4.7 \pm 1.1$ & $10.3 \pm 2.3$ & $2.2\times$** & 0.87 \\
    Java   & $4.6 \pm 1.2$ & $10.2 \pm 3.2$ & $2.2\times$** & 0.86 \\
    Python & $5.1 \pm 1.2$ & $10.5 \pm 3.5$ & $2.1\times$** & 0.85 \\
    \bottomrule
    \multicolumn{5}{l}{\footnotesize Significance: * $p < 0.05$, ** $p < 0.01$.} \\
\end{tabular}
\end{table}

The structural refactoring improvements come at an increased computational overhead, summarized in Table~\ref{tab:tokens}. The Basic system prompt contained $3{,}495$ tokens. Including the MCP server tool definitions raised it to $6{,}370$ tokens. Per task, the Guided runs consume $\sim$$3.0$--$5.4\times$ the total tokens of the Basic runs. The increase is dominated by input tokens ($\sim$$98$\% of the Guided total), which inflate as the multi-turn feedback loop stacks prior context in the agent's conversation.

Output token generation increases by $\sim$$2\times$, and interaction turns rise from $\sim$$5$ to $\sim$$10$ per task. Wilcoxon signed-rank tests show that all increases are significant ($p < 0.01$) with large effect sizes ($r = 0.71$--$0.87$). The agent actively uses the iterative feedback loop, repeatedly querying, patching, and re-querying, which produces the improved refactoring outcomes of \RQ{1}--\RQ{2} at the cost of more tokens, context-window usage, and turns.

\finding{5}{Metric guidance costs $3.0$--$5.4\times$ the total tokens and $2.1$--$2.2\times$ the interaction turns per task ($p < 0.01$, large effects), driven primarily by context accumulated over the iterative feedback loop.}

\subsection{Observations from Logs}
\label{disc:logs}

Under the Basic prompt, the sampled agent typically follows a linear trajectory. It reads the file, identifies maintainability issues from pre-trained knowledge (no external tools), applies a single patch (often dominated by renaming), and terminates. The Guided agent instead treats \texttt{code\_health\_review} as an active feedback loop, proposing a structural change, applying it, re-querying, and iterating. In one Java session (\texttt{java\_guided1}), the agent decomposed a large function into several helper methods and then, after the tool still reported a residual code smell, initiated further decomposition round.

CH does not increase monotonically after each iteration. Edits are applied before their effect on CH is known, an intermediate refactoring can lower the score before a later one recovers it. In \texttt{python\_guided2}, the first structural edit dropped CH from $8.95$ to $7.83$, introduced two complexity smells, before a second edit lifted it to $9.24$. In \texttt{java\_guided1}, an intermediate rewrite reduced CH from $8.28$ to $7.90$ before returning to $8.28$ in a subsequent edit.

The loop can also spend several iterations on a file with little room for improvement (ceiling effect). In \texttt{java\_guided2}, a file that already scored a healthy $9.68$, the agent rewrote the file multiple times, which failed, without moving the score, then stopped, reasoning:

\begin{quote}
\textit{``The code health score is still 9.68 and the Primitive Obsession issue persists. However, I've already made the most meaningful refactor I can [\ldots] the main algorithmic logic is actually quite clean and well-structured. The `Primitive Obsession' issue is inherent to the nature of the problem being solved.''}
\end{quote}

The sampled logs also show the agent reasoning about the limits of metric compliance. In \texttt{java\_guided3}, after three failed attempts to remediate a ``Bumpy Road Ahead'' code smell inside a depth-first search function, the agent concluded:

\begin{quote}
\textit{``The remaining `Bumpy Road Ahead' issue in the dfs method appears to be inherent to the algorithm's nature --- it naturally has two distinct logical sections that are difficult to separate without changing the algorithm itself. This is a common pattern in graph algorithms\ldots\ Since the task specifically asks to follow CodeScene's code health metric without altering semantic behavior\ldots\ I'll consider this refactoring complete.''}
\end{quote}

In \texttt{python\_guided1}, the agent attempted to verify its generated refactoring, by executing the refactored program on a sample input passed through a mocked \texttt{stdin}. The verification crashed because of issues in the mocked test, and the agent fell back to a simpler syntax check (\texttt{python3 -m py\_compile}) before terminating. In the other observed logs, the sessions terminated without compilation or running the code.

\section{Discussion}
\label{sec:discussion}

Our findings reveal tensions in autonomous code maintenance. Agents can appear active without improving structure, and structural code transformations raise the risk of breaking behavior while increasing token usage. This section interprets the results and derives implications.

\subsection{Towards Structural Remediation}

When prompted generically to ``improve maintainability'', our agent, powered by a capable open-weight model gravitates toward low-risk, cosmetic edits. Under the Basic prompt, variable renaming alone accounted for $41$--$57$\% of detected refactoring operations. These edits usually preserve behavior (higher pass rates) but barely move internal code quality, with fewer than half of Basic runs improving CH at all ($31$--$47$\%). This is the same consistency-oriented profile observed in public agentic commits of GitHub pull requests~\citep{horikawa2025agenticrefactoringempiricalstudy, ottenhof2026agentsrefactorempiricalstudy}, and it can create an illusion of maintenance, where plausible-looking refactoring commits mostly leave TD in place.

Explicit metric feedback changes this profile. \textit{Extract Method} grows into the most frequent refactoring operation in Java and the second most frequent in Python, suggesting that the capability for structural refactoring is present in the model but remains latent without concrete, actionable targets. The MCP server translates the abstract instruction of ``improve maintainability'' into explicit directives (e.g., a reported \textit{Complex Conditional} code smell with its location). This supports and extends the argument of \citet{cordeiroLLMdrivenCodeRefactoring2025}, that explicit purpose aligns AI refactorings with intent of uplifting code quality. Delivered iteratively through a tool, it can change what kind of edits the agent attempts.

\subsection{Refactoring Side Effects}

The side effects of structural remediation differed in magnitude across languages. In Java, the generated refactorings led to a net introduction of \textit{Excess Arguments} ($1{,}080$ to $4{,}825$), \textit{Primitive Obsession} ($2{,}274$ to $6{,}078$), \textit{String-Heavy Arguments} ($39$ to $138$), and \textit{Constructor Over-Injection} ($23$ to $198$). When decomposing methods, our agent defaulted to procedural parameter passing instead of synthesizing appropriate data transfer objects or managing class-level state, trading a complexity smell for a coupling smell. \textit{Primitive Obsession} is also the only tested smell that the Basic agent remediates more often than the Guided agent (\RQ{2}). Overall, the Basic agent's refactorings show the same side effects patterns, but at a smaller magnitude (Figure~\ref{fig:smells}). Consequently, Guidance amplifies this rather than introducing it. C++ and Python show the same direction with smaller increases in \textit{Excess Arguments} ($-65.8$\% and $-57.9$\% versus $-346.8$\% in Java), and C++ also gains \textit{Primitive Obsession} instances ($2{,}839$ to $5{,}057$). In our setup, the agent dismantles poor structures when directed to them but lacks the architectural foresight to recompose the parts into clean encapsulated abstractions.

This aligns with~\citet{Wanatabe2026}, that even when agent generated pull requests are functionally correct, reviewers frequently intervene to perform additional restructuring. Our results further indicate this pattern, where a reviewer would be presented with a Guided refactoring that is successfully decomposed, but with the introduced smells relating to coupling, requiring cognitive attention in spotting and remediating suboptimally designed dataflow between components.

Increased compile and test failures is the other side effect. Compile errors would be caught by an agent that compiles its code before terminating, and part of the runtime and test failures by executing and testing its generated code. Our agent rarely verified its own edits in the small sample of conversation logs (Section~\ref{disc:logs}). The prompts do not instruct the agents to verify its refactorings. For any such verification step to occur in our setup, it depends on the LLM and the harness to trigger it. Different LLMs, harness or instructions could compile and test its refactoring before terminating more frequently, and this would avoid most of those fault classes.

\subsection{AI-Friendliness Threshold}

The moderation analysis (\RQ{3}) supports the AI-Friendliness hypothesis~\citep{thoughtworks2025, borg2026codemachinesjusthumans} in the agentic setting. For refactorings on solutions with high CH, guidance encourages structural transformations largely without regression in pass rate (with -3 pp for Java and Python and +8 pp for C++ at the model-implied point of CH 9).

On degraded solutions, guidance still lifts CH more than the Basic prompt but also breaks behavior more often. The Guided advantage in effective refactoring rate falls from $+26$--$34$ pp at CH 9 to $+1$, $-8$, and $-25$ pp at CH 5 in C++, Java, and Python. We attribute this to structural edits being harder for the LLM to perform correctly on low-quality code (AI-Friendliness).
However, Guided refactorings that pass tests in low CH are still more impactful in maintainability gain (Section~\ref{sec:rq3}). We therefore read the region below the break-even points as an indication of where the agent needs additional support in performing structural transformations correctly.

The model-implied break-even points ($\text{CH}^* \approx 4.9$--$7.1$) should be treated as rough estimates specific to our languages, LLM, harness, and metric rather than universal constants. The low-CH bins are sparse, and per-bin deltas are noisy. However, the consistent interaction across languages and refactoring outcomes supports the qualitative conclusion that agents are better suited to maintaining moderately healthy code than improving severely degraded code. For practice, this suggests a staging strategy, where human developers (or more conservative tooling) may first need to lift files above the AI-Friendliness threshold before delegating autonomous remediation, for optimal results.

\subsection{Optimizing the Metric}

CH serves as both the guidance signal and the outcome measure in our design. The gains of \RQ{1} therefore show that the agent can effectively optimize a target metric, not that CH improvements are necessarily synonymous with maintainability in some tool-agnostic way. Optimizing a proxy invites the typical risks of measure gaming~\citep{Thomas2022}. Three observations mitigate this concern. First, CH as a metric is independently validated against external outcomes such as defect density and development velocity~\citep{tornhillCodeRedBusiness2022, borgIncreasingNotDiminishing2024}, making it a suitable and verified proxy for maintainability. Second, the behavioral shift is corroborated by RefactoringMiner, an independent tool that detects the structural operations from the ASTs without reference to CH. Third, the smell-level analysis (Figure~\ref{fig:smells} and Table~\ref{tab:mcnemar_tests}) exposes where optimization produced regressions (new coupling smells) rather than hiding them in an aggregate score. Still, part of the measured benefit is inevitably an artifact of evaluating with the guiding metric. Consequently, results should be read as improvements under the CH metric, which we accept by design. The object of study is not whether CH is the right maintainability target, but whether an agent can move it on its own, and where that ability breaks down. Reading the results as improvements under the CH metric is therefore the intended interpretation. The points at which the agent stalls, halts short of a perfect score, or trades one smell for another are as much a part of the contribution as the aggregate gains.

The conversation logs (Section~\ref{disc:logs}) illustrate two such points. First, the agent does not raise CH monotonically, as an intermediate edit can lower the score before a subsequent edit recovers it. A run that terminates during such a dip ends with a file worse than it started, which is one possible reason that some of the refactorings end below their initial CH (\RQ{1}). Second, the agent halts instead of reaching a perfect CH score of $10.0$ when it judges a remaining smell inherent to the algorithm, prioritizing behavior preservation and existing algorithm design over continuously iterating and remediating all CH reported code smells.

\subsection{Overhead of Guided Refactoring}
\label{disc:overhead}

The token overhead ($\sim$$3$--$5\times$ increase) frames metric guidance as an investment decision. The iterative read $\to$ query $\to$ plan $\to$ patch $\to$ re-query loop expands the context window on every turn, with input tokens dominating this growth. The conversation logs (Section~\ref{disc:logs}) show that this loop can continue when the metric offers no actionable opportunity (for the specific agent), where a file was rewritten multiple times without CH improvement. Practitioners must weigh these computational costs against the anticipated maintainability gains. For a healthy file containing a localized smell, guided remediation is more safe, but may also risk minimal or no gain in quality due to a ceiling effect. Conversely, na\"{\i}vely deploying autonomous agents on severely degraded code is both more expensive and more prone to behavioral regressions (\RQ{3}).

Framed per unit of success rather than per run, the overhead is smaller. Dividing mean tokens by the effective-refactoring rate, the Guided experiments spend $\sim$$470$k, $357$k, and $268$k tokens per effective refactoring in C++, Java, and Python, against $\sim$$235$k, $110$k, and $159$k for Basic. The cost per effective refactoring is therefore $2.0\times$ (C++), $3.2\times$ (Java), and $1.7\times$ (Python) instead.

This risk-cost profile suggests a metric-routed maintenance strategy for batch scenarios. Practitioners can use the initial CH as a routing signal, delegating healthy files to autonomous guided agents while flagging degraded files for human-supervised refactoring. A retrospective, in-sample simulation of this policy on our dataset, routing solutions to be refactored to the Guided prompt only when its initial CH is at or above the language-specific break-even point $\text{CH} \ge \text{CH}^*$ (Figure~\ref{fig:rq3}), achieves the same effective refactoring rate as an all-Guided baseline while improving both pass rates and token efficiency. Because $\text{CH}^*$ is estimated from the same runs, we present this as an illustrative policy rather than a validated router. For example, in Java, this routed policy yields a $54.9$\% effective rate and a $69.0$\% pass rate using $194$k mean tokens, outperforming the all-Guided baseline ($54.8$\% effective, $68.6$\% pass, $198$k tokens). C++ and Python show the same pattern.

While these simulated performance margins are narrow, due to our majority healthy corpus ($98$--$99$\% of refactorings route to Guided), the protective value of this policy scales with the proportion of TD in a codebase, since it avoids correctness risk on the degraded files. Routing a degraded file to the Basic agent increases the chance of a conservative shallow refactoring, trading the larger improvement potential that guidance delivers when it succeeds. In practice, such a potential threshold is specific to the setting (harness, model, codebase) and would need to be calibrated locally.

\subsection{Implications}

\textbf{For practitioners.} Do not equate agent refactoring activity with TD remediation, as unguided agents produce plausible but superficial edits. Exposing quality gates to the agent's loop via MCP changed outcomes in our study, but deployment should be gated on initial code quality, and practitioners should expect token usage increase of several times that of unguided runs.

\textbf{For tool builders.} The trade-off smells of \RQ{2} suggest guidance should be asymmetric. Tools should warn about the smells a pending edit could potentially introduce (e.g., excessive parameters after an extraction), not just report the smells already present. The failure shifts in \RQ{4} suggest agents would benefit from integration-level verification (compilation checks, self-generated smoke tests) after structural edits, rather than syntactic validation alone.

\textbf{For researchers.} The observed interaction between guidance and initial code quality demonstrates that average treatment effects are insufficient for evaluating agentic interventions. Moderation analyses can establish the boundary conditions under which such tools shift from beneficial to detrimental. Additionally, our findings motivate multi-metric evaluation frameworks that isolate the guiding signal from the assessment instruments, disentangling proxy optimization from maintainability gains.

\section{Threats to Validity}
\label{sec:threats}

\subsection{Construct Validity}

The internal code quality measure is CodeScene's CH, which is also the metric provided to the Guided agent. Improvements in effective refactoring rate can therefore be interpreted as improvements under the CH metric rather than a general notion of maintainability.

Throughout the study, token counts are approximated. \texttt{gptme} reports them with \texttt{tiktoken} (\texttt{cl100k\_base}) instead of Qwen3-Coder's own tokenizer, and therefore, the absolute counts may differ from what the LLM processed and generated. The same measurement tools and tokenizers are applied identically across both experimental conditions, relative comparisons between runs remain consistent.

\subsection{Internal Validity}

Each task was executed once per condition without resampling (temperature 0.7). Stochastic variability is mitigated by our large dataset size rather than by repeated refactoring runs per solution. Guided treatment acts as a bundle (instruction, tool availability, and tool-based feedback), and this design cannot isolate the observed effects of guidance with parts of its components, such as what the tools report (smells and CH) or differences in the instructions (Appendix~\ref{sec:prompt-templates}).

The $674$ Guided runs where the MCP server process failed before any refactoring was produced, are counted as failed Guided runs. These cases are skewed towards larger C++ files with lower initial CH (median $181$ vs.\ $127$ lines, and $13.1$\% of C++ Guided runs below CH 7 versus $3.3$\% above), consequently, this contributes to the lower Guided pass rate on degraded files (\RQ{3}).

The 20-minute execution timeout risks categorizing slow but correct refactorings as failures, though this is unlikely given the small size of the solutions. Additionally, the context window limit (32k tokens) penalizes long multi-turn sessions. SLoC is relatively flat across the CH range (Table~\ref{tab:loc_bins}), which rules out file length as a driver of the primary \RQ{3} interaction. However, low-CH code, contains more code smells, and is therefore more entangled, and Guided sessions on it consume more tokens. Consequently, context pressure may contribute to parts of the lower Guided pass rate on degraded files.

\subsection{External Validity}

The dataset relies on isolated, single-file competitive programming solutions. While this enables our strict paired design and behavioral oracles, it omits the cross-file dependency resolution, build systems, and retrieval logic required in enterprise maintenance. Consequently, while the mechanisms we observe (e.g., shifts toward structural edits, generation of localized coupling smells) align with studies of agents on production repositories~\citep{horikawa2025agenticrefactoringempiricalstudy, ottenhof2026agentsrefactorempiricalstudy}, the success rates and break-even thresholds will likely differ in multi-module codebases.

Finally, results are tied to a specific agent harness (\texttt{gptme}), one open-weight model (Qwen3-Coder-30B-A3B), and one maintainability metric (CH). More capable models or harnesses may shift the break-even thresholds, although we expect the trends and the moderation by initial code quality to hold (reflection for headroom of improvement). Replication with other agents, frontier models, independent maintainability measures, multi-file production software projects, and a broader range of smell types is needed before the trends can be generalized.

\section{Conclusion}
\label{sec:conclusion}

We presented a paired empirical study of $49{,}240$ agentic refactoring runs across C++, Java, and Python, evaluating what happens when an autonomous coding agent is given iterative access to static-analysis feedback. By connecting \texttt{gptme} to the CodeScene MCP server, our findings challenge the assumption that agents natively understand how to reduce TD. Instead, integrating these agents into software maintenance requires navigating four distinct trade-offs:

\begin{itemize}
    \item \textbf{Structural refactorings.} Without explicit targets, agents default to cosmetic, consistency-oriented edits. Actionable feedback successfully redirects them toward structural transformations (e.g., \textit{Extract Method}), raising the share of runs that improve CH by $34$--$39$ pp and the effective refactoring rate by $21$--$26$ points under the guiding metric.
    \item \textbf{The AI-Friendliness threshold.} Metric guidance does not offer a universal remedy for severe TD. The agent's ability to safely restructure code depends on the initial quality of that code. Below language-specific break-even points (CH $\approx 4.9$--$7.1$), pushing the agent to decompose tangled logic increases the rate of behavioral regressions.
    \item \textbf{Shifting the failure risk.} When agents undertake deeper structural edits, their failure modes change. The failure mix shifts away from primitive crashes (most prominently C++ segmentation faults) toward subtle logic errors and timeouts. This shift highlights the importance of comprehensive test suites, as the faults introduced by guided agents are more often semantic rather than syntactic.
    \item \textbf{Computational verhead.} Iterating against a static-analysis tool requires more compute. The continuous feedback loop doubles the interaction turns of the agent and consumes $\sim$$3$--$5\times$ more tokens than unguided refactoring.
\end{itemize}

These findings outline several directions for future research. First, ablation studies should decompose the guidance bundle, isolating the value of the metric semantics from the general benefit of any external feedback tool. Second, maintenance tools should monitor trade-off smells, warning the agent about new architectural issues introduced during refactorings rather than only reporting existing smells to remove. Third, multi-metric designs that use separate instruments for guidance and assessment would ensure agents improve underlying maintainability rather than optimize a specific proxy. Finally, scaling this experimental design to multi-file, enterprise repositories will clarify how autonomous AI teammates navigate cross-module dependencies and real-world environments.

\backmatter

\bmhead{Acknowledgements}

This work was partially supported by the Wallenberg AI, Autonomous Systems and Software Program WASP (WARA-Ops), funded by the Knut and Alice Wallenberg Foundation, and partly by the Competence Centre NextG2Com funded by the VINNOVA program for Advanced Digitalisation with grant number 2023-00541.

\section*{Declarations}

\bmhead{Funding}
See Acknowledgements.

\bmhead{Conflict of interest}
Markus Borg is employed by CodeScene, the vendor of the CodeScene CLI tool whose CH metric and MCP server are studied in this work.

\bmhead{Ethics approval and consent to participate}
Not applicable.

\bmhead{Consent for publication}
Not applicable.

\bmhead{Data availability}
The full dataset of agent runs, including per-task outcomes, CH scores, RefactoringMiner output, and sampled conversation logs, is available in the replication package~\citep{replication}.

\bmhead{Materials availability}
Not applicable.

\bmhead{Code availability}
All analysis scripts and prompt templates are available in the replication package~\citep{replication}.

\bmhead{Author contribution}
\textbf{Nadim Hagatulah}: Conceptualization, Methodology, Software, Investigation, Formal analysis, Data curation, Writing -- original draft. \textbf{Markus Borg}: Conceptualization, Supervision, Visualization, Writing -- review \& editing. \textbf{Emma S\"{o}derberg}: Conceptualization, Supervision, Writing -- review \& editing.

\begin{appendices}

\section{Prompt Templates}
\label{sec:prompt-templates}

\subsection{Basic Prompt}
\begin{small}
\begin{verbatim}
# Purpose
Act as an expert software engineer specialized in {language}
code.

# Task
Refactor the code in `{filename}` to improve maintainability and code quality
without altering semantic behaviour.

# Requirements
Strictly follow these requirements:
- Use the `save` tool to overwrite `{filename}` with the full refactored
  content, do not introduce any other new files.
- Do not edit the `{filename}` name.
- Do not alter the program's logic or output behavior.
- Do not introduce new libraries, tools, packages or dependencies.
- Do not perform general fixes or modifications, outside of remediating the
  maintainability issues.

# Execution
Now proceed with the refactoring of `{filename}` by:
1. Using the `read` tool to understand the content of `{filename}`.
2. Review and analyze the code and identify maintainability
   issues.
3. Plan refactoring steps to remediate the identified issues.
4. Execute the refactoring steps and overwrite `{filename}` with the full
   refactored program using the `save` tool.
5. Use the `read` tool to understand the new content of `{filename}`.
6. Validate and analyze that all maintainability issues are remediated and no
   new ones are found.
7. If new issues are found, go back to step 3, otherwise
   terminate.
\end{verbatim}
\end{small}

\subsection{Guided Prompt}
\begin{small}
\begin{verbatim}
# Purpose
Act as an expert software engineer specialized in {language}
code.

# Task
Refactor the code in `{filename}` to improve maintainability and code quality,
guided strictly by CodeScene's code health metric, without altering semantic
behavior.

# Tools
The CodeScene MCP server is available as codescene. Use:
- code_health_review -- analyze code health for a specific file.

# Requirements
Strictly follow these requirements:
- Use the `save` tool to overwrite `{filename}` with the full refactored
  content, do not introduce any other new files.
- Do not edit the `{filename}` name.
- Do not alter the program's logic or output behavior.
- Do not introduce, import, download, or install any new libraries, packages,
  tools, or dependencies. Use only standard libraries or existing imports.
- Do not make any changes, except those strictly necessary to remediate the
  identified maintainability issues.

# Execution
Now proceed with the refactoring of `{filename}` by:
1. Using the `read` tool to understand the content of `{filename}`.
2. Running code_health_review to identify concrete
   maintainability issues.
3. Planning refactoring steps to remediate the identified issues.
4. Executing the refactoring steps and overwriting `{filename}` with the full
   refactored program using the `save` tool.
5. Running code_health_review on the updated `{filename}`.
6. Validating that all identified maintainability issues are remediated and no
   new ones are introduced.
7. If new issues are found, go back to step 3, otherwise
   terminate.
\end{verbatim}
\end{small}

\section{GEE Model Specification}
\label{sec:gee-details}

The moderation analysis of \RQ{3} uses a single GEE logistic regression, fit separately to three binary outcomes of run $i$: whether it is an effective refactoring, whether it strictly increases CH, and whether it passes the tests. Writing $Y_i$ for whichever outcome is modeled,
\begin{equation}
    \label{eq:gee}
    \operatorname{logit}\Pr(Y_i = 1) = \beta_{0}
    + \beta_{\mathrm{P}}\, G_i
    + \beta_{\mathrm{CH}}\, H_i
    + \beta_{\mathrm{P\times CH}}\, G_i H_i,
\end{equation}
where $G_i \in \{0, 1\}$ indicates the Guided prompt and $H_i$ is the initial CH of the file. The interaction $\beta_{\mathrm{P\times CH}}$ is the moderation term: its sign says whether guidance helps more ($>0$) or less ($<0$) as initial CH rises, and Guided and Basic predictions coincide at the break-even point $\text{CH}^* = -\beta_{\mathrm{P}} / \beta_{\mathrm{P\times CH}}$. Each model uses an exchangeable working correlation and is clustered by task. Table~\ref{tab:gee_regression} lists the coefficients for all three outcomes, matching the three rows of Figure~\ref{fig:rq3}.

\begin{table}[htbp]
    \centering
    \caption{GEE regressions, following Equation~\ref{eq:gee}: an effective refactoring, a strict CH improvement, and a passing test run, matching the three rows of Figure~\ref{fig:rq3}. Coefficients are on the log-odds scale with standard errors in parentheses, clustered by task. $\text{CH}^*$ is the break-even initial CH where the Guided and Basic predictions coincide; it reads $>$$10$ or $<$$1$ when the arms do not cross within the CH scale, and 95\% CIs are given for the primary effective-refactoring outcome.}
    \label{tab:gee_regression}
    \footnotesize
    \setlength{\tabcolsep}{3pt}
    \begin{tabular}{lrrrr}
        \toprule
        \textbf{Term} & \textbf{Overall}            & \textbf{C++}                & \textbf{Java}               & \textbf{Python}             \\
                      & \scriptsize{($N=49{,}240$)} & \scriptsize{($N=19{,}650$)} & \scriptsize{($N=19{,}590$)} & \scriptsize{($N=10{,}000$)} \\
        \midrule
        \multicolumn{5}{@{}l}{\textbf{Effective refactoring}} \\
        Intercept ($\beta_{0}$)
                           & $0.934~(0.095)$**  & $1.203~(0.158)$**  & $3.121~(0.172)$**  & $3.695~(0.424)$**  \\
        Guided at CH $=0$ ($\beta_{\mathrm{P}}$)
                           & $-1.721~(0.115)$** & $-2.229~(0.188)$** & $-2.437~(0.203)$** & $-3.922~(0.512)$** \\
        Initial CH, Basic slope ($\beta_{\mathrm{CH}}$)
                           & $-0.236~(0.012)$** & $-0.361~(0.021)$** & $-0.476~(0.021)$** & $-0.483~(0.047)$** \\
        Moderation ($\beta_{\mathrm{P\times CH}}$)
                           & $0.338~(0.014)$**  & $0.455~(0.024)$**  & $0.415~(0.025)$**  & $0.554~(0.057)$**  \\
        Break-even $\text{CH}^*$
                           & --- & $4.9~[4.6, 5.2]$ & $5.9~[5.6, 6.2]$ & $7.1~[6.7, 7.5]$ \\
        \midrule
        \multicolumn{5}{@{}l}{\textbf{Increasing CH}} \\
        Intercept ($\beta_{0}$)
                           & $2.922~(0.096)$**  & $3.061~(0.148)$**  & $5.413~(0.180)$**  & $5.978~(0.447)$**  \\
        Guided at CH $=0$ ($\beta_{\mathrm{P}}$)
                           & $0.567~(0.140)$**  & $-0.850~(0.186)$** & $2.487~(0.320)$**  & $0.080~(0.702)$\phantom{**}  \\
        Initial CH, Basic slope ($\beta_{\mathrm{CH}}$)
                           & $-0.413~(0.012)$** & $-0.507~(0.019)$** & $-0.688~(0.022)$** & $-0.694~(0.049)$** \\
        Moderation ($\beta_{\mathrm{P\times CH}}$)
                           & $0.130~(0.017)$**  & $0.331~(0.024)$**  & $-0.079~(0.037)$*  & $0.162~(0.077)$*   \\
        Break-even $\text{CH}^*$
                           & --- & $2.6$ & $>$$10$ & $<$$1$ \\
        \midrule
        \multicolumn{5}{@{}l}{\textbf{Passing the tests}} \\
        Intercept ($\beta_{0}$)
                           & $-0.372~(0.097)$** & $0.390~(0.129)$**  & $0.128~(0.177)$\phantom{**}  & $-0.257~(0.478)$\phantom{**} \\
        Guided at CH $=0$ ($\beta_{\mathrm{P}}$)
                           & $-2.477~(0.127)$** & $-2.974~(0.180)$** & $-2.421~(0.221)$** & $-3.538~(0.580)$** \\
        Initial CH, Basic slope ($\beta_{\mathrm{CH}}$)
                           & $0.164~(0.012)$**  & $0.013~(0.016)$\phantom{**}  & $0.136~(0.022)$**  & $0.200~(0.053)$**  \\
        Moderation ($\beta_{\mathrm{P\times CH}}$)
                           & $0.278~(0.016)$**  & $0.371~(0.023)$**  & $0.249~(0.028)$**  & $0.370~(0.065)$**  \\
        Break-even $\text{CH}^*$
                           & --- & $8.0$ & $9.7$ & $9.6$ \\
        \bottomrule
        \multicolumn{5}{@{}l}{\footnotesize Significance: * $p < 0.05$, ** $p < 0.01$.}
    \end{tabular}
\end{table}

\section{Refactoring Operation Levels}
\label{sec:operation-levels}

Table~\ref{tab:operation_levels} lists the $84$ RefactoringMiner operation types detected in the Java and Python runs, grouped by the refactoring level of \citet{horikawa2025agenticrefactoringempiricalstudy}.

\begin{table}[htbp]
    \centering
    \caption{Refactoring operation types detected in the Java and Python runs, by refactoring level.}
    \label{tab:operation_levels}
    \footnotesize
    \begin{tabular}{@{} l p{10.6cm} @{}}
        \toprule
        \textbf{Level} & \textbf{Operation types} \\
        \midrule
        High ($42$)   & Add Attribute Annotation, Add Attribute Modifier, Add Class Modifier, Add Method Annotation, Add Method Modifier, Add Parameter, Add Thrown Exception Type, Change Attribute Access Modifier, Change Class Access Modifier, Change Method Access Modifier, Change Thrown Exception Type, Change Type Declaration Kind, Encapsulate Attribute, Extract Superclass, Merge Attribute, Merge Class, Merge Parameter, Modify Method Annotation, Move And Rename Attribute, Move And Rename Class, Move And Rename Method, Move Attribute, Move Class, Move Method, Parameterize Variable, Pull Up Attribute, Pull Up Method, Push Down Method, Remove Attribute Annotation, Remove Attribute Modifier, Remove Class Annotation, Remove Class Modifier, Remove Method Annotation, Remove Method Modifier, Remove Parameter, Remove Thrown Exception Type, Rename Attribute, Rename Class, Rename Method, Reorder Parameter, Split Attribute, Split Parameter \\
        \midrule
        Medium ($18$) & Change Attribute Type, Change Parameter Type, Change Return Type, Extract And Move Method, Extract Attribute, Extract Class, Extract Method, Extract Subclass, Inline Attribute, Inline Method, Localize Parameter, Merge Method, Move And Inline Method, Parameterize Attribute, Replace Anonymous With Class, Replace Attribute With Variable, Replace Variable With Attribute, Split Method \\
        \midrule
        Low ($24$)    & Add Variable Annotation, Add Variable Modifier, Change Variable Type, Extract Variable, Inline Variable, Invert Condition, Merge Catch, Merge Conditional, Merge Variable, Move Code, Remove Parameter Modifier, Remove Variable Annotation, Remove Variable Modifier, Rename Parameter, Rename Variable, Replace Anonymous With Lambda, Replace Attribute, Replace Conditional With Ternary, Replace Generic With Diamond, Replace Loop With Pipeline, Replace Pipeline With Loop, Split Conditional, Split Variable, Try With Resources \\
        \bottomrule
    \end{tabular}
\end{table}

\section{SLoC across CH}
\label{sec:sloc}

The Table~\ref{tab:loc_bins} shows SLoC across CH bins.

\begin{table}[htbp]
    \centering
    \caption{SLoC of the original file by initial CH bin (CH floored to an integer).}
    \label{tab:loc_bins}
    \footnotesize
    \setlength{\tabcolsep}{5pt}
    \begin{tabular}{@{} c cc cc cc @{}}
        \toprule
        & \multicolumn{2}{c}{\textbf{C++}} & \multicolumn{2}{c}{\textbf{Java}} & \multicolumn{2}{c}{\textbf{Python}} \\
        \cmidrule(lr){2-3} \cmidrule(lr){4-5} \cmidrule(lr){6-7}
        \textbf{CH} & \textbf{Mean $\pm$ SD} & \textbf{Median} & \textbf{Mean $\pm$ SD} & \textbf{Median} & \textbf{Mean $\pm$ SD} & \textbf{Median} \\
        \midrule
        1  & $132 \pm 36$ & $119$ & $177 \pm 27$ & $172$ & $92$          & $92$  \\
        2  & $120 \pm 39$ & $112$ & $116 \pm 53$ & $94$  & --            & --    \\
        3  & $114 \pm 38$ & $107$ & $167 \pm 52$ & $170$ & $73$          & $73$  \\
        4  & $127 \pm 36$ & $123$ & $149 \pm 41$ & $142$ & $101 \pm 30$  & $101$ \\
        5  & $133 \pm 41$ & $129$ & $158 \pm 45$ & $158$ & $133 \pm 55$  & $128$ \\
        6  & $141 \pm 38$ & $143$ & $166 \pm 44$ & $170$ & $105 \pm 28$  & $100$ \\
        7  & $139 \pm 35$ & $138$ & $170 \pm 41$ & $173$ & $113 \pm 34$  & $107$ \\
        8  & $130 \pm 26$ & $123$ & $169 \pm 36$ & $167$ & $110 \pm 45$  & $101$ \\
        9  & $129 \pm 26$ & $122$ & $163 \pm 35$ & $158$ & $107 \pm 36$  & $99$  \\
        10 & --           & --    & --           & --    & $88 \pm 19$   & $81$  \\
        \bottomrule
    \end{tabular}
\end{table}

\section{Code Smell Glossary}
\label{sec:smell-glossary}

Table~\ref{tab:smell_glossary} summarizes the CodeScene code smells referenced in this study. Definitions paraphrase the CodeScene documentation\footnote{\url{https://codescene.io/docs}}, and the final column lists the detection threshold used, which are the default thresholds. Table~\ref{tab:smell_counts} reports the smell instances before and after refactorings.

\begin{table}[htbp]
    \centering
    \caption{CodeScene code smells referenced in this paper, with the tool's default detection threshold. Thresholds are CodeScene's default configuration and are identical across the C++, Java, and Python rule sets. ``--'' marks smells flagged by a composite rule (Brain Method) or a rule match with no tunable numeric threshold (Global Conditionals, String-Heavy Arguments).}
    \label{tab:smell_glossary}
    \footnotesize
    \setlength{\tabcolsep}{4pt}
    \begin{tabular}{@{} l p{5.6cm} p{2.7cm} @{}}
        \toprule
        \textbf{Code Smell} & \textbf{Description} & \textbf{Threshold} \\
        \midrule
        Large Method & A function whose length exceeds the tool's size threshold. & $\ge 70$ lines \\
        Brain Method & A long function that also concentrates high complexity and deep nesting, centralizing much of the file's logic. & -- \\
        Complex Method & A function whose cyclomatic complexity exceeds the threshold. & CC $\ge 9$ \\
        Complex Conditional & A branch condition combining multiple logical operators. & $\ge 2$ operators \\
        Deep, Nested Complexity & Control structures nested beyond the depth threshold. & Depth $\ge 4$ \\
        Bumpy Road Ahead & A function containing multiple consecutive chunks of nested conditional logic, indicating missing abstractions. & $\ge 2$ bumps \\
        Global Conditionals & Conditional logic at module scope, outside any function. & -- \\
        Overall Code Complexity & The file-level average cyclomatic complexity across its functions. & Mean CC $\ge 4$ \\
        Code Duplication & Duplicated logic across functions in the file. & $\ge 10$ lines, $\ge 75\%$ similar \\
        Excess Arguments & A function taking more arguments than the threshold. & $> 4$ args \\
        Primitive Obsession & Interfaces that pass many primitive-typed values where domain abstractions would be expected. & $\ge 30\%$ primitives \\
        String-Heavy Arguments & A function signature dominated by string-typed arguments. & -- \\
        Constructor Over-Injection & A constructor receiving an excessive number of dependencies or arguments. & $> 5$ args \\
        \bottomrule
    \end{tabular}
\end{table}

\begin{table}[htbp]
    \centering
    \caption{Smell categories before and after refactoring, with net reduction $= 100 \times (\text{Before} - \text{After}) / \text{Before}$, over the tasks with a CodeScene review of both refactored files (Figure~\ref{fig:smells}). Before counts the initial files, which are shared by both runs. Smells with fewer than 20 instances before refactoring in a language are omitted.}
    \label{tab:smell_counts}
    \footnotesize
    \setlength{\tabcolsep}{4pt}
    \begin{tabular}{@{} l l r rr rr @{}}
        \toprule
        & & & \multicolumn{2}{c}{\textbf{Basic}} & \multicolumn{2}{c}{\textbf{Guided}} \\
        \cmidrule(lr){4-5} \cmidrule(lr){6-7}
        \textbf{Language} & \textbf{Code Smell} & \textbf{Before} & \textbf{After} & \textbf{Net (\%)} & \textbf{After} & \textbf{Net (\%)} \\
        \midrule
        \multirow{11}{*}{\textbf{C++}}
            & Large Method             & $3{,}092$ & $2{,}913$ & $+5.8$ & $308$ & $+90.0$ \\
            & Brain Method             & $50$ & $71$ & $-42.0$ & $23$ & $+54.0$ \\
            & Complex Conditional      & $14{,}364$ & $11{,}651$ & $+18.9$ & $6{,}949$ & $+51.6$ \\
            & Complex Method           & $11{,}148$ & $11{,}447$ & $-2.7$ & $9{,}261$ & $+16.9$ \\
            & Deep, Nested Complexity  & $5{,}670$ & $6{,}357$ & $-12.1$ & $4{,}494$ & $+20.7$ \\
            & Bumpy Road Ahead         & $11{,}038$ & $12{,}359$ & $-12.0$ & $10{,}745$ & $+2.7$ \\
            & Overall Code Complexity  & $2{,}709$ & $2{,}835$ & $-4.7$ & $3{,}137$ & $-15.8$ \\
            & Code Duplication         & $2{,}253$ & $2{,}299$ & $-2.0$ & $3{,}614$ & $-60.4$ \\
            & Excess Arguments         & $2{,}983$ & $3{,}674$ & $-23.2$ & $4{,}946$ & $-65.8$ \\
            & Primitive Obsession      & $2{,}839$ & $2{,}933$ & $-3.3$ & $5{,}057$ & $-78.1$ \\
            & String-Heavy Arguments   & $40$ & $48$ & $-20.0$ & $148$ & $-270.0$ \\
        \midrule
        \multirow{12}{*}{\textbf{Java}}
            & Large Method             & $2{,}018$ & $925$ & $+54.2$ & $78$ & $+96.1$ \\
            & Brain Method             & $23$ & $24$ & $-4.3$ & $6$ & $+73.9$ \\
            & Complex Conditional      & $10{,}825$ & $7{,}960$ & $+26.5$ & $4{,}246$ & $+60.8$ \\
            & Complex Method           & $11{,}414$ & $10{,}462$ & $+8.3$ & $5{,}876$ & $+48.5$ \\
            & Deep, Nested Complexity  & $5{,}635$ & $4{,}883$ & $+13.3$ & $2{,}410$ & $+57.2$ \\
            & Bumpy Road Ahead         & $9{,}441$ & $9{,}316$ & $+1.3$ & $5{,}784$ & $+38.7$ \\
            & Overall Code Complexity  & $4{,}155$ & $3{,}087$ & $+25.7$ & $1{,}456$ & $+65.0$ \\
            & Code Duplication         & $7{,}857$ & $8{,}505$ & $-8.2$ & $7{,}529$ & $+4.2$ \\
            & Excess Arguments         & $1{,}080$ & $2{,}823$ & $-161.4$ & $4{,}825$ & $-346.8$ \\
            & Primitive Obsession      & $2{,}274$ & $3{,}398$ & $-49.4$ & $6{,}078$ & $-167.3$ \\
            & String-Heavy Arguments   & $39$ & $60$ & $-53.8$ & $138$ & $-253.8$ \\
            & Constructor Over-Injection & $23$ & $28$ & $-21.7$ & $198$ & $-760.9$ \\
        \midrule
        \multirow{9}{*}{\textbf{Python}}
            & Large Method             & $172$ & $160$ & $+7.0$ & $47$ & $+72.7$ \\
            & Global Conditionals      & $646$ & $176$ & $+72.8$ & $249$ & $+61.5$ \\
            & Complex Conditional      & $1{,}269$ & $1{,}122$ & $+11.6$ & $597$ & $+53.0$ \\
            & Complex Method           & $3{,}544$ & $3{,}329$ & $+6.1$ & $1{,}494$ & $+57.8$ \\
            & Deep, Nested Complexity  & $2{,}419$ & $2{,}074$ & $+14.3$ & $1{,}099$ & $+54.6$ \\
            & Bumpy Road Ahead         & $4{,}887$ & $4{,}539$ & $+7.1$ & $2{,}908$ & $+40.5$ \\
            & Overall Code Complexity  & $308$ & $368$ & $-19.5$ & $332$ & $-7.8$ \\
            & Code Duplication         & $290$ & $705$ & $-143.1$ & $519$ & $-79.0$ \\
            & Excess Arguments         & $717$ & $829$ & $-15.6$ & $1{,}132$ & $-57.9$ \\
        \bottomrule
    \end{tabular}
\end{table}

\end{appendices}

\clearpage
\bibliography{sn-bibliography}

\end{document}